# Self-Consistent Field Theory for Semiflexible Gaussian Chain Model

Yutaka Oya [1, *] and Toshihiro Kawakatsu [2]

[1] *Department of Material Science and Technology, Graduate School of Advanced Engineering, Tokyo University of Science, Tokyo 125-8585, Japan.*
[2] *Department of Physics, Graduate School of Science, Tohoku University, Sendai 980-8578, Japan.*

## Abstract

Self-consistent field theory (SCFT) is one of the useful methods to simulate phase separated structures of multi-component polymer systems. In this article, we propose an SCFT for semiflexible polymer melts, where the basic equations for the SCFT are derived by introducing a bending stiffness into a flexible Gaussian bead-spring model and taking its continuous limit. Our SCFT is described by a coupled modified diffusion equations for the statistical weight of the chain conformation (path integral), which is a perturbation to semiflexible chains from the flexible Gaussian chain model. Using our modified diffusion equations, we investigated the influences of the bending stiffness on the conformations of symmetric semiflexible diblock copolymer in a strongly segregated lamellar structures and on the order-disorder transition.

# I. Introduction

Micro-phase separations of polymers, which are ubiquitous phenomena in both biological systems and industrial products, have been investigated using various computer simulation approaches [1]. Field-theoretic approaches (FTA) are especially useful in studying domain morphologies on mesoscopic scales [2-5].

Previous studies have elucidated various phase-separated structures in the melt state of block polymers with various chain configurations, including linear [4, 6], comb [7, 8], star [9, 10], and ring polymers [11, 12]. The efficiency of the artificial control of diverse phase-separated structures has been demonstrated through confinement in rigid narrow spaces [13-16] or the incorporation of nanoparticles [17-19]. By replicating equilibrium structures arising from interactions between bio-membranes and phase-separated structures [20-23], FTA has extracted deeper understanding of metabolic processes, such as endocytosis and exocytosis. These equilibrium structures can be quantitatively determined from various candidates by comparing their free energies. Furthermore, the method can be extended to non-equilibrium dynamics by combining Ginzburg-Landau theory with energy functions (Onsager variational principle) [24-26], leading to the studies on the steady states of flowing vesicles [27] and domain morphology grown in reactive polymer systems [28]. Therefore, FTA enables us to derive equilibrium and steady-state phase diagrams from weak to strong segregation regimes based on the free energy and the dissipation functions, which usually cannot easily be evaluated by particles-based approaches such as coarse-grained molecular dynamics simulations or dissipative particle dynamics simulations [29].

Among these FTAs, self-consistent field theory (SCFT) is the one that can reproduce experimental results quantitatively. This is because SCFT is capable of accounting for all possible conformations of polymer chains with any topology under an external field [2-5]. Many previous studies using SCFT have successfully reproduced a wide variety of experimental observations by assuming the polymer chain to be flexible. On the other hand, there have been relatively few studies that focus on the stiffness of polymer chains compared to those focusing on flexible polymer chains [30-44]. In this article, we apply SCFT to study the effect of bending stiffness of polymers on phase separation.

In the standard SCFT, the treatment of the statistical weight of the chain conformations, which is called a "path-integral", is different between a flexible chain and a semiflexible chain. For the flexible chain, we use a continuous limit of a flexible bead-spring model, where the bond connecting adjacent segments along the polymer chain is modelled by a harmonic spring [5, 45]. As the statistical distribution of such polymer chain is Gaussian, we hereafter call this model as "flexible Gaussian bead-spring model". On the other hand, the semiflexible polymer chain is usually described by the so-called worm-like chain model [5, 45], where the polymer is modelled by a chain made of rigid rods with bending stiffness between adjacent rods. This worm-like chain model has been used to simulate phase separation of polymers with large persistent length such as DNA, proteins, rod-like viruses and actin filaments [46].

In both flexible Gaussian bead-spring model and the worm-like chain model, the statistical unit, i.e. the bead or the rod, is called a statistical segment, which is a coarse-grained object composed of several monomers so that we can neglect the detailed molecular structure except for the connection between adjacent segments. Due to the different treatments of the statistical segment between the flexible Gaussian bead-spring model and the worm-like chain model, the form of the partial differential equation (so-called "modified diffusion equation") for the path-integral is also different for these two types of polymer models. In the flexible Gaussian bead-spring model, the Hamiltonian of a polymer chain $H$ is defined as

$$H = \frac{3k_{\mathrm{B}}T}{2b_0^2}\sum_{i=0}^{N-1}|\boldsymbol{u}_i|^2 = \frac{3k_{\mathrm{B}}T}{2b_0^2\Delta i}\sum_{i=0}^{N-1}|\boldsymbol{u}_i|^2\Delta i \cong \frac{3k_{\mathrm{B}}T}{2b^2}\int_0^N |\boldsymbol{u}(s)|^2 ds \tag{1}$$

where $\boldsymbol{u}_i$ is the bond vector defined by $\boldsymbol{u}_i \equiv \boldsymbol{r}_{i+1} - \boldsymbol{r}_i$, $\boldsymbol{r}_i$ being the position of the $i$-th segment. The Kuhn length of the segment, $b_0$, is redefined by $b^2 \equiv b_0^2\Delta i$. In the final expression of eq.(1), we used the continuum limit, where discrete index $i$ is replaced by a continuous variable $s$. In such a continuous limit, the range of segment index $i$ that runs over $i = 0,1,\cdots,N$ corresponds to $0 \leq s \leq N$. Then, we obtain the diffusion equation for the path-integral as

$$\frac{\partial Q(s,\boldsymbol{r})}{\partial s}=\frac{b^2}{6}\nabla^2 Q(s,\boldsymbol{r}), \tag{2}$$

where $Q(s,\boldsymbol{r})$ is the path-integral of a chain whose $s$-th segment is located at position $\boldsymbol{r}$ assuming that end segment $s$ = 0 can be anywhere in the system. On the other hand, in the worm-like chain model, Hamiltonian of a polymer chain and the diffusion equation for the path-integral in the continuous limit are described by

$$H=\frac{\kappa_0}{2}\sum_{i=0}^{N-1}|\boldsymbol{c}_i|^2, \tag{3}$$

$$\frac{\partial Q(s,\boldsymbol{r},\hat{\boldsymbol{u}})}{\partial s}+\hat{\boldsymbol{u}}(s)\cdot\boldsymbol{\nabla}Q(s,\boldsymbol{r},\hat{\boldsymbol{u}})=\frac{1}{2\kappa}\nabla_{\hat{u}}{}^2 Q(s,\boldsymbol{r},\hat{\boldsymbol{u}}), \tag{4}$$

where $\boldsymbol{c}_i$ is the curvature vector at $i$-th segment defined by $\boldsymbol{c}_i \equiv \hat{\boldsymbol{u}}_{i+1}-\hat{\boldsymbol{u}}_i$, $\hat{\boldsymbol{u}}_i$ being the unit bond vector of the $i$-th segment, and $\nabla_{\hat{\boldsymbol{u}}}$ is partial derivative with respect to $\hat{\boldsymbol{u}}$. $\kappa_0$ is bending elasticity constant and $\kappa \equiv \kappa_0/\Delta i$. $Q(s,\boldsymbol{r},\hat{\boldsymbol{u}})$ is the path-integral of a polymer chain whose $s$-th segment is located at position $\boldsymbol{r}$ and connected to ($s$+1)-th segments by bond vector $\hat{\boldsymbol{u}}$. Phase separations of biomolecules have been reproduced by solving eq. (4) in both real space and wave number space [30-44]. These studies have shown that the bending stiffness has a significant effect on the conformations of the polymers, leading to a change in the phase transition lines in the phase diagram compared with that obtained with the flexible Gaussian bead-spring model [44]. Note that the unit bond vector $\hat{\boldsymbol{u}}$ is introduced as an additional variable to the path-integral as shown in eq. (4), which requires a much larger calculation cost than that of flexible Gaussian bead-spring model.

As is obvious from the above introduction of the flexible Gaussian bead-spring model for a flexible polymer chain and the worm-like chain model for a semiflexible polymer chain, these two models are based on completely different descriptions of the bonds. In the flexible Gaussian bead-spring model, the bonds are described by harmonic springs with fluctuating bond lengths and a vanishing persistence length. This is a result of coarse-graining procedure, which maps a realistic polymer chain into a chain of flexible bead-springs when the length scale of the coarse-grained bond becomes larger than the persistence length of the original realistic polymer chain. On the other hand, in the worm-like chain model, the degree of coarse-graining on bonds is smaller than the persistence length, leading to an essentially fixed bond length. Thus, the worm-like chain model can be regarded as a perturbation to the semiflexible chain region from the completely rigid polymer chain model that has an infinite persistence length.

In the present paper, we develop a model for semiflexible polymer chains based on a perturbation from the opposite limit of the polymer chain from the completely rigid chain model. We introduce a bending stiffness into the flexible Gaussian bead-spring model as a perturbation. This corresponds to a situation where the coarse-graining procedure on a realistic polymer chain is terminated while the persistence length is finite. This assumption leads to the following Hamiltonian of a single semiflexible polymer chain:

$$H_0 = \frac{3k_\mathrm{B}T}{2b_0^2}\sum_{i=0}^{N-1}|\boldsymbol{u}_i|^2 - \frac{k_0}{2}\sum_{i=0}^{N-2}\boldsymbol{u}_i\cdot\boldsymbol{u}_{i+1}, \tag{5}$$

where $k_0$ is the bending stiffness constant. The second term on the right-hand side of eq. (5) represents the orientational correlations between neighboring bonds, which leads to a similar bending stiffness as the worm-like chain model (eq. (3)). The main difference between our bending stiffness term (i.e. the $2^{nd}$ term on the right-hand side of eq.(5)) and the corresponding energy in the worm-like chain model is the fact that our model is described using the bond vector $\boldsymbol{u}$, whose length is flexible, instead of the unit bond vector $\hat{\boldsymbol{u}}$ in the worm-like chain model. This assumption is crucial to keep our chain statistics within the Gaussian statistics. Because of the second term of eq. (5), bond vectors are no longer statistically independent, and the corresponding modified diffusion equation for SCFT calculations should be reconstructed.

Hereafter, we refer to our polymer model that incorporates stiffness through perturbation from the Gaussian chain as the "semiflexible Gaussian chain model". It should be noted that a similar polymer chain model was first introduced by Harris and Hearst in 1966 [47], and previous studies have investigated the statistical properties of this polymer chain model [48-52]. However, to the best of our knowledge, this is the first report proposing an SCFT framework based on semiflexible Gaussian chain model. In this study, we investigate the effect of the bending stiffness on the phase separation by solving the modified diffusion equation corresponding to the Hamiltonian expressed by eq. (5). As a target of this study, we choose microphase separations of symmetric semiflexible diblock-copolymer melts.

The remainder of this paper is organized as follows. The next section is devoted to the derivation of the modified diffusion equation of our semiflexible polymer model. In the third section, the simulation results and discussions about diblock co-polymers that are composed of symmetric semiflexible/semiflexible blocks are presented. Finally, we conclude our results in the final section and suggest some future directions.

# II. Simulation methods

## 1. Derivation of the modified diffusion equation.

In this section, we derive a set of modified diffusion equations for a semiflexible polymer chain in a melt using an incompressible homopolymer melt as an example. First, we define the Hamiltonian based on a discrete beads-spring model, and then we move to a continuous description where the polymer chain is represented as a continuous string.

Let us consider a melt of semiflexible polymers, where each polymer chain is composed of $N$+1 segments connected by harmonic bonds with the root mean square bond length $b_0$ (i.e. Kuhn statistical length). We assume a bending stiffness between adjacent bonds with a bending stiffness modulus $k_0$. Within the mean field approximation, the non-bonded pair interactions between segments are replaced by their interactions with a mean field $V(\boldsymbol{r})$. Then, the Hamiltonian of a single polymer chain in the melt is given by

$$H = H_0 + \sum_{i=0}^{N} V(\boldsymbol{r}_i), \tag{6}$$

$$H_0 = \frac{3k_\mathrm{B}T}{2b_0^2}\sum_{i=0}^{N-1}|\boldsymbol{u}_i|^2 - \frac{k_0}{2}\sum_{i=0}^{N-2}\boldsymbol{u}_i\cdot\boldsymbol{u}_{i+1}, \tag{7}$$

where $\boldsymbol{r}_i$ is the position of $i$-th segment and $\boldsymbol{u}_i \equiv \boldsymbol{r}_{i+1} - \boldsymbol{r}_i$ is the $i$-th bond vector. Using the Hamiltonian, eqs. (6) and (7), we define the path-integral of this chain under the constraint that the 0-th segment (one end segment) is located at $\boldsymbol{r}_0$ and the other end segment, $N$-th segment, is located at $\boldsymbol{r}_N$ as follows:

$$Q(0,\boldsymbol{r}_0\,;N,\boldsymbol{r}_N) \equiv \frac{1}{Z(1)}\int d\Gamma' \exp\left\{-\frac{1}{k_\mathrm{B}T}H_0 - \frac{1}{k_\mathrm{B}T}\left(\frac{1}{2}V(\boldsymbol{r}_0) + \sum_{i=1}^{N-1}V(\boldsymbol{r}_i) + \frac{1}{2}V(\boldsymbol{r}_N)\right)\right\}, \tag{8}$$

$$Z(1) \equiv \int d\Gamma \exp\left\{-\frac{1}{k_\mathrm{B}T}H_0\right\}, \tag{9}$$

where $Z(1)$ is the partition function of a single ideal polymer chain without the effect of the mean field $V$. The integral $\int\{*\}d\Gamma \equiv \int d\boldsymbol{r}_1 \cdots \int d\boldsymbol{r}_N\{*\}$ represents integration of $\{*\}$ over the entire configuration space of segment positions $\{\boldsymbol{r}_i\}$, and $\int\{*\}d\Gamma'$ represents a similar integration but with constraints on the positions of two end segments $(0, \boldsymbol{r}_0)$ and $(N, \boldsymbol{r}_N)$. In eq. (8), the contribution from the mean field $V(\boldsymbol{r})$ is halved for end segments.

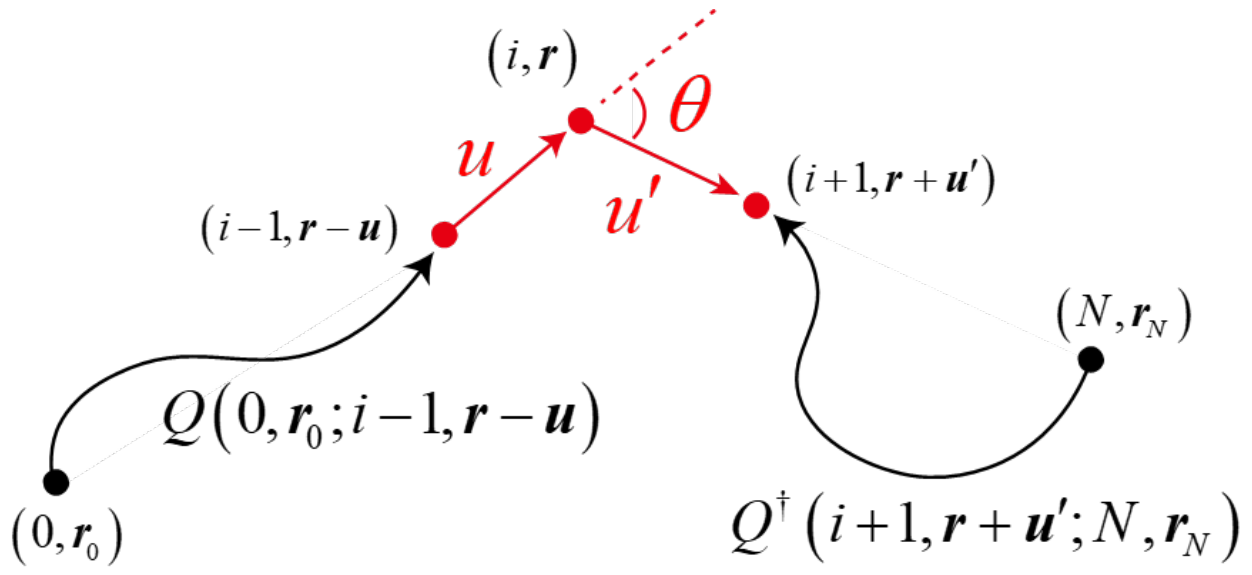

Figure 1. Schematic illustration of state variables of segments of a single semiflexible polymer chain which is divided into three sub-chains. Each state variable is represented in a form of (segment index, position of the segment), e.g., $(0, \boldsymbol{r}_0)$ means that 0-th segment is located at position $\boldsymbol{r}_0$. The bond vector $\boldsymbol{u}$ connects $(i$-1)-th segment and $i$-th segment, and $\boldsymbol{u}'$ connects $i$-th segment and $(i$+1)-th segment. $\theta$ is the angle formed by bond vectors $\boldsymbol{u}$ and $\boldsymbol{u}'$.

The same is true for the end segments of individual sub-chains. This is necessary to guarantee the symmetry between the two end segments of a sub-chain when we cut a chain into sub-chains.

To explain how to derive the modified diffusion equation for the path-integral, in Fig.1 we give an illustration of the state variables of a single polymer chain whose two ends are located at $\boldsymbol{r}_0$ and $\boldsymbol{r}_N$, respectively. In this Figure, two types of path-integrals, i.e., a path-integral in the "forward direction" $Q$ and a path-integral in the "backward-direction" $Q^\dagger$ are defined. For the path-integral in the forward direction $Q$, the starting and end segments are defined as the 0-th and $N$-th segments, which are located at $\boldsymbol{r}_0$ and $\boldsymbol{r}_N$, respectively. On the other hand, for the path-integral in the backward direction $Q^\dagger$, the starting and end segments are defined as the $N$-th and 0-th segments in an opposite manner to $Q$. As is shown in Fig. 1, a polymer chain is divided into three sub-chains. The first sub-chain has its end segments at $(0, \boldsymbol{r}_0)$ and $(i - 1, \boldsymbol{r} - \boldsymbol{u})$. The second one has its end segments at $(i -1, \boldsymbol{r} - \boldsymbol{u})$ and $(i + 1, \boldsymbol{r} + \boldsymbol{u}')$ (second sub-chain is highlighted by red in Fig.1), and the third one has its end segments at $(i + 1, \boldsymbol{r} + \boldsymbol{u}')$ and $(N, \boldsymbol{r}_N)$. Sum of the statistical weights of all possible conformations of the first sub-chain is given by $Q(0, \boldsymbol{r}_0; i - 1, \boldsymbol{r} - \boldsymbol{u})$, and that of the third sub-chain is given by $Q^\dagger(i+1, \boldsymbol{r} + \boldsymbol{u}'; N, \boldsymbol{r}_N)$. The product of these two path-integrals for the first and the third sub-chains and the Boltzmann factor of the second sub-chain gives the sum of the statistical weights of the whole polymer chain for all possible conformations under the constraint that the 0-th, $i$-th, and $N$-th segments are located at positions $\boldsymbol{r}_0$, $\boldsymbol{r}$, and $\boldsymbol{r}_N$. Then, we obtain

$$Q(0,\boldsymbol{r}_0;N,\boldsymbol{r}_N)=\frac{1}{Z(1)}\int d\boldsymbol{r}\int d\boldsymbol{u}\int d\boldsymbol{u}'\ Q(0,\boldsymbol{r}_0;i-1,\boldsymbol{r}-\boldsymbol{u})Q^{\dagger}(i+1,\boldsymbol{r}+\boldsymbol{u}';0,\boldsymbol{r}_N)$$
$$\times\exp\left[-\frac{1}{k_{\mathrm{B}}T}\left\{h_0(\boldsymbol{u},\boldsymbol{u}')-\frac{1}{2}\{V(\boldsymbol{r}-\boldsymbol{u})+2V(\boldsymbol{r})+V(\boldsymbol{r}+\boldsymbol{u}')\}\right\}\right], \tag{10}$$

$$h_0(\boldsymbol{u},\boldsymbol{u}')\equiv\frac{3k_{\mathrm{B}}T}{2b_0^2}|\boldsymbol{u}|^2+\frac{3k_{\mathrm{B}}T}{2b_0^2}|\boldsymbol{u}'|^2-\frac{k_0}{2}\boldsymbol{u}\cdot\boldsymbol{u}' \tag{11}$$

where the exponential factor represents the Boltzmann factor of the second sub-chain composed of three segments represented by red part in Fig. 1, and $h_0$ is the Hamiltonian of the second sub-chain without non-bond interactions.

The calculation cost for the path-integrals can largely be reduced by introducing path-integrals that are integrated over the positions of their end segments. Let us introduce notations as $\int Q(0,\boldsymbol{r}_0;i,\boldsymbol{r})d\boldsymbol{r}_0\equiv Q(i,\boldsymbol{r})$ and $\int Q^{\dagger}(i,\boldsymbol{r};N,\boldsymbol{r}_{\mathrm{N}})d\boldsymbol{r}_N\equiv Q^{\dagger}(i,\boldsymbol{r})$. We perform Taylor series expansions of $Q$, $Q^{\dagger}$ and $V$ in the right-hand side of eq. (10) with respect to $\boldsymbol{u}$ and $\boldsymbol{u}$' around the position $\boldsymbol{r}$. As a result, the right-hand side of eq. (10) can be rewritten in the following form:

$$\frac{1}{Z(1)}\int d\boldsymbol{r}\int d\boldsymbol{u}\int d\boldsymbol{u}'\left\{Q(i-1,\boldsymbol{r})-\nabla Q(i-1,\boldsymbol{r})\cdot\boldsymbol{u}+\frac{1}{2}\nabla\nabla Q(i-1,\boldsymbol{r}):\boldsymbol{u}\boldsymbol{u}\right\}$$
$$\times\left\{Q^{\dagger}(i+1,\boldsymbol{r})+\nabla Q^{\dagger}(i+1,\boldsymbol{r})\cdot\boldsymbol{u}'+\frac{1}{2}\nabla\nabla Q^{\dagger}(i+1,\boldsymbol{r}):\boldsymbol{u}'\boldsymbol{u}'\right\}$$
$$\times\exp\left\{-\frac{2}{k_{\mathrm{B}}T}V(\boldsymbol{r})\right\}\exp\left\{-\frac{1}{k_{\mathrm{B}}T}h_0(\boldsymbol{u},\boldsymbol{u}')\right\}. \tag{12}$$

Let us consider the factor $\exp\left\{-\frac{1}{k_{\mathrm{B}}T}h_0(\boldsymbol{u}_1,\boldsymbol{u}_2)\right\}$ in eq. (12). We diagonalize $h_0$ by introducing the variable $\boldsymbol{v}$ and $\boldsymbol{v}'$ defined by

$$\begin{pmatrix}\boldsymbol{v}\\ \boldsymbol{v}'\end{pmatrix}\equiv\frac{1}{\sqrt{2}}\begin{pmatrix}\boldsymbol{u}+\boldsymbol{u}'\\ \boldsymbol{u}-\boldsymbol{u}'\end{pmatrix}, \tag{13}$$

where $\boldsymbol{v}$ represents coarse-grained bond vector consisting of two adjacent bond vectors, and $\boldsymbol{v}'$ is regarded as the curvature vector that is the difference between two adjacent bond vectors. Then, eq. (11) is rewritten in terms of $\boldsymbol{v}$ and $\boldsymbol{v}'$ as

$$h_0'(\boldsymbol{v},\boldsymbol{v}') \equiv \frac{k_{\mathrm{B}}T}{2}\left(\frac{3}{b_0^2}-k_0\right)|\boldsymbol{v}|^2 + \frac{k_{\mathrm{B}}T}{2}\left(\frac{3}{b_0^2}+k_0\right)|\boldsymbol{v}'|^2. \tag{14}$$

Substituting eq. (14) into eq. (10) and integrating it over $\boldsymbol{v}$ and $\boldsymbol{v}'$, we obtain the following expression:

$$\begin{aligned}\int d\boldsymbol{r}_0 \int d\boldsymbol{r}_N\ Q(0,\boldsymbol{r}_0\,; N,\boldsymbol{r}_N) = \frac{1}{Z(1)}\int d\boldsymbol{r}\Bigg[\left(1-\frac{2}{k_{\mathrm{B}}T}V(\boldsymbol{r})\right)Q(i-1,\boldsymbol{r})Q^{\dagger}(i+1,\boldsymbol{r})\\ +\frac{1}{4}\left(\frac{b_0^2}{3-k_0b_0^2}+\frac{b_0^2}{3+k_0b_0^2}\right)\{\nabla^2 Q(i-1,\boldsymbol{r})Q^{\dagger}(i+1,\boldsymbol{r})+Q(i-1,\boldsymbol{r})\nabla^2 Q^{\dagger}(Ni+1,\boldsymbol{r})\}\\ +\frac{1}{2}\left(\frac{b_0^2}{3-k_0b_0^2}-\frac{b_0^2}{3+k_0b_0^2}\right)\nabla Q(i-1,\boldsymbol{r})\cdot\nabla Q^{\dagger}(i+1,\boldsymbol{r})+o(\nabla V(\boldsymbol{r})b)\Bigg],\end{aligned} \tag{15}$$

where $o(\nabla V(\boldsymbol{r})b_0)$ indicates terms of order $\nabla V(\boldsymbol{r})$ multiplied by the Kuhn length $b_0$ and higher order terms.

Now, we move to a continuous description of the polymer chain contour, where the discrete segment index $i$ is replaced by a continuous variable $s$, and accordingly, we replace $i \pm \Delta i \equiv i \pm 1$ by $s \pm ds$. In this case, the square of the effective bond length $b_0^2 \times \Delta i \equiv b_0^2 \times 1$ should be replaced by $b^2\, ds$ so that the end-to-end distance is unchanged when we change the mesh size $ds$.

The continuous model can be obtained by taking the limit of $ds \to 0$. To obtain the continuous model, we expand the integrand in the right-hand side of eq. (15) in Taylor series in $ds$ around the position $s$ along the chain contour and retain terms up to 1st order in $ds$, which gives

$$
\begin{aligned}
&\int d\boldsymbol{r}_0 \int d\boldsymbol{r}_N \; Q(0,\boldsymbol{r}_0\,;N,\boldsymbol{r}_N\,) = \frac{1}{Z(1)}\int d\boldsymbol{r}\Big[\, Q(s,\boldsymbol{r})Q^{\dagger}(s,\boldsymbol{r}) \\
&+ds\left\{\left(-\frac{\partial Q(s,\boldsymbol{r})}{\partial s}Q^{\dagger}(s,\boldsymbol{r}) + Q(s,\boldsymbol{r})\frac{Q^{\dagger}(s,\boldsymbol{r})}{\partial s}\right) - \frac{2}{k_{\mathrm{B}}T}V(\boldsymbol{r})Q(s,\boldsymbol{r})Q^{\dagger}(s,\boldsymbol{r})\right. \\
&+\frac{1}{4}\left(\frac{b^2}{3-k_0b^2ds}+\frac{b^2}{3+k_0b^2ds}\right)\left\{\nabla^2 Q(s,\boldsymbol{r})Q^{\dagger}(s,\boldsymbol{r}) + Q(s,\boldsymbol{r})\nabla^2 Q^{\dagger}(s,\boldsymbol{r})\right\} \\
&\left.-\frac{1}{2}\left(\frac{b^2}{3-k_0b^2ds}-\frac{b^2}{3+k_0b^2ds}\right)\nabla Q(s,\boldsymbol{r})\cdot\nabla Q^{\dagger}(s,\boldsymbol{r})\right\}\Big].
\end{aligned}
\tag{16}
$$

Equation (16) gives an extension of Chapman-Kolmogorov relation for the partition function of a single chain, i.e. $\int d\boldsymbol{r}_0 \int d\boldsymbol{r}_{\mathrm{N}}\, Q(0,\boldsymbol{r}_0;\mathrm{N},\boldsymbol{r}_{\mathrm{N}})$, up to the 1st order in $ds$ except for the $k_0b^2$ part that apparently produces terms of $o(ds^2)$. This relation, eq.(16), should be satisfied for any small $ds$. Equating the 0th order terms in $ds$ of the both sides of eq.(16) gives the Chapman-Kolmogorov relation for a flexible chain without bending stiffness. On the other hand, equating the 1st order terms on both sides gives the modified diffusion equation for the path integrals $Q(\mathrm{s},\boldsymbol{r})$ and $Q^{\dagger}(\mathrm{s},\boldsymbol{r})$.

Here, there are two ways of transferring to the continuous model by taking the limit of $ds \to 0$. One is to keep $k_0$ constant, and the other is to assume the persistence length of the chain constant. In the former procedure, the 1st order terms of eq.(16) reduce to the usual modified diffusion equation for an ideal flexible Gaussian chain without bending stiffness (i.e. Markovian chain). On the other hand, in the latter procedure, to keep the persistence length constant, we should assume $k_0 \equiv \frac{k}{ds}$ with $k$ constant when $ds \to 0$. In this case, equating the 1st order terms in $ds$ in the both sides of eq.(16)

becomes

$$\left(-\frac{\partial Q(s,\boldsymbol{r})}{\partial s}Q^{\dagger}(s,\boldsymbol{r})+Q(s,\boldsymbol{r})\frac{Q^{\dagger}(s,\boldsymbol{r})}{\partial s}\right)-\frac{2}{k_{\mathrm{B}}T}V(\boldsymbol{r})Q(s,\boldsymbol{r})Q^{\dagger}(s,\boldsymbol{r})$$
$$+\frac{1}{4}\left(\frac{b^2}{3-kb^2}+\frac{b^2}{3+kb^2}\right)\{\nabla^2 Q(s,\boldsymbol{r})Q^{\dagger}(s,\boldsymbol{r})+Q(s,\boldsymbol{r})\nabla^2 Q^{\dagger}(s,\boldsymbol{r})\} \tag{17}$$
$$\left.-\frac{1}{2}\left(\frac{b^2}{3-kb^2}-\frac{b^2}{3+kb^2}\right)\nabla Q(s,\boldsymbol{r})\cdot\nabla Q^{\dagger}(s,\boldsymbol{r})\right]=0.$$

This equation is our basic equation for calculating the path-integrals of a semiflexible polymer in its melt state.

As the two path-integrals $Q$ and $Q^{\dagger}$ are coupled in eq. (17), it is rather easier to solve the following set of two coupled evolution equations for $Q$ and $Q^{\dagger}$ rather than to solve eq. (17) directly.

$$\frac{\partial\, Q(s,\boldsymbol{r})}{\partial s}=\left[g_{+}(k,b)\nabla^2-g_{-}(k,b)\frac{\nabla Q^{\dagger}(s,\boldsymbol{r})}{Q^{\dagger}(s,\boldsymbol{r})}\cdot\nabla-\frac{1}{k_{\mathrm{B}}T}V(\boldsymbol{r})\right]Q(s,\boldsymbol{r}), \tag{18}$$

$$\frac{\partial\, Q^{\dagger}(s,\boldsymbol{r})}{\partial s}=\left[-g_{+}(k,b)\nabla^2+g_{-}(k,b)\frac{\nabla Q(s,\boldsymbol{r})}{Q(s,\boldsymbol{r})}\cdot\nabla+\frac{1}{k_{\mathrm{B}}T}V(\boldsymbol{r})\right]Q^{\dagger}(s,\boldsymbol{r}), \tag{19}$$

$$g_{\pm}(k,b)\equiv\frac{1}{4}\left(\frac{b^2}{3-kb^2}\pm\frac{b^2}{3+kb^2}\right), \tag{20}$$

which are sufficient condition for eq. (17). When $k=0$, the differential operators of these equations reduce to those of the modified diffusion equations for standard flexible Gaussian chain model given in eq. (2). Note that the standard modified diffusion equations for the flexible Gaussian bead-spring model (eq.(2)) and the worm-like chain model (eq.(4)) have Markovian forms. Especially, in the worm-like chain model, the evolution equation for the path-integral is transformed into a Markovian form by introducing an extra variable $\boldsymbol{u}$ into the definition of $Q$ and $Q^{\dagger}$ to make adjacent bond doublets to be statistically independent [30-44]. On the other hand, our equations, eqs. (18) and (19), are not Markovian as we do not introduce the extra variable to $Q$ and $Q^{\dagger}$. Instead, we have to solve the set of eqs. (18) and (19) iteratively. This is extremely efficient in reducing the computer memory in performing simulations on higher dimensional systems.

Let us consider the physical meaning of the coefficients $g_{+}(k,b)$ and $g_{-}(k,b)$ defined in eq. (20). We define the statistical average of a quantity $\{*\}$ over the canonical ensemble for the Hamiltonian of a single ideal polymer chain with the bending stiffness energy between two consecutive bonds $h_0'$ in eq. (14), as follows:

$$\langle * \rangle_0 = \int d\boldsymbol{v} \int d\boldsymbol{v}' * \exp\left\{-\frac{1}{k_{\mathrm{B}}T} h_0'(\boldsymbol{v},\boldsymbol{v}')\right\} / \int d\boldsymbol{v} \int d\boldsymbol{v}' \exp\left\{-\frac{1}{k_{\mathrm{B}}T} h_0'(\boldsymbol{v},\boldsymbol{v}')\right\}. \tag{21}$$

Using eqs. (13) and (21), the statistical average of the square of a bond vector and that of the inner product of neighboring bond vectors are obtained as

$$\begin{aligned} \langle |\boldsymbol{u}|^2 \rangle_0 &= \frac{1}{4} \langle |\boldsymbol{v}|^2 + |\boldsymbol{v}'|^2 \rangle_0 = 6 g_{+}(k,b), \\ \langle \boldsymbol{u}(s) \cdot \boldsymbol{u}(s+ds) \rangle_0 &= \frac{1}{4} \langle |\boldsymbol{v}|^2 - |\boldsymbol{v}'|^2 \rangle_0 = 6 g_{-}(k,b). \end{aligned} \tag{22}$$

From eq. (22), we understand that $g_{+}(k,b)$ and $g_{-}(k,b)$ represent the statistical average of square of length of a single bond and that of the correlation between adjacent bonds. These averages are functions of the bending stiffness $k$ and Kuhn length $b$ of the chain.

Here, we consider the relationship between semiflexible Gaussian chain model introduced in this study and the standard worm-like chain model. From eq. (20), we find that $g_{\pm}(k, b)$ diverges when $k$ approaches either $-3/b^2$ or $3/b$.[2] In particular, when $k$ approaches $3/b^2$, the Hamiltonian represented by eqs. (6) and (7) reduces to the Hamiltonian of the worm-like chain model because in such a limit the sum of the energy of the harmonic spring and that of the directional correlation is regarded as the bending stiffness energy. Note that semiflexible Gaussian chain model allows the negative value of the bending stiffness. This negative bending stiffness has been used in the molecular dynamics simulation with Kremer-Grest model [53], while it cannot be used in the standard worm like chain model for numerical stability reasons. In the standard worm-like chain model, the statistical average of the angle between neighboring bonds is quantitatively controlled by the parameter $\langle \cos\theta \rangle_0 \equiv \langle \hat{\boldsymbol{u}}(s) \cdot \hat{\boldsymbol{u}}(s + ds) \rangle_0$, where $\hat{\boldsymbol{u}}$ is the unit bond vector. On the other hand, in the present study, this parameter is approximated by

$$\langle \cos\theta \rangle_0^{\mathrm{eff}} \equiv \frac{\langle \boldsymbol{u}(s) \cdot \boldsymbol{u}(s + ds) \rangle_0}{\langle |\boldsymbol{u}|^2 \rangle_0} = \frac{kb^2}{3}. \tag{23}$$

This parameter $\langle \cos\theta \rangle_0^{\mathrm{eff}}$ can take a value between -1 and 1 as is shown in APPENDIX A. Using eqs. (22) and (23), the modified diffusion equations represented by eqs. (18) and (19) are rewritten as

$$\frac{\partial\, Q(s, \boldsymbol{r})}{\partial s} = \frac{1}{6} \langle |\boldsymbol{u}|^2 \rangle_0 \left[ \nabla^2 - \langle \cos\theta \rangle_0^{\mathrm{eff}} \frac{\nabla Q^{\dagger}(s, \boldsymbol{r})}{Q^{\dagger}(s, \boldsymbol{r})} \cdot \nabla \right] Q(s, \boldsymbol{r}) - \frac{1}{k_{\mathrm{B}} T} V(\boldsymbol{r}) Q(s, \boldsymbol{r}), \tag{24}$$

$$\frac{\partial Q^{\dagger}(s,\boldsymbol{r})}{\partial s} = -\frac{1}{6}\langle|\boldsymbol{u}|^2\rangle_0\left[\nabla^2 - \langle\cos\theta\rangle_0^{\mathrm{eff}}\frac{\nabla Q(s,\boldsymbol{r})}{Q(s,\boldsymbol{r})}\cdot\nabla\right]Q^{\dagger}(s,\boldsymbol{r}) + \frac{1}{k_{\mathrm{B}}T}V(\boldsymbol{r})Q^{\dagger}(s,\boldsymbol{r}) \quad (25)$$

Equations (24) and (25) are our final forms of the modified diffusion equations for semiflexible Gaussian polymer chains. In the following, we use eqs. (24) and (25) to simulate micro-phase separation of semiflexible Gaussian polymer melts.

## 2. SCFT calculation and simulation conditions

In this section, we turn our attention to micro/macro phase separations of semiflexible block copolymer melts/semiflexible polymer blends. To describe these inhomogeneous systems, we introduce a closed set of equations for the SCFT calculation. Denoting the polymer chain species by the index $\alpha$ while introducing another index $K = A, B, \cdots$ for the segment species, total free energy of the system ($F$) is given by [5, 45]

$$F = -\sum_{\alpha} M_{\alpha}\log\int d\boldsymbol{r}\, Q_{\alpha}(s,\boldsymbol{r})Q_{\alpha}{}^{\dagger}(s,\boldsymbol{r}) - \sum_{K}\int d\boldsymbol{r}\phi_K(\boldsymbol{r})V_K(\boldsymbol{r}) + E(\{\phi_K(\boldsymbol{r})\}), \quad (26)$$

$$E(\{\phi_K(\boldsymbol{r})\}) = \frac{1}{2}\sum_{K,K'}\int d\boldsymbol{r}\,\chi_{KK'}\phi_K(\boldsymbol{r})\phi_{K'}(\boldsymbol{r}) + \int d\boldsymbol{r}\,\gamma(\boldsymbol{r})\left[\sum_{K}\phi_K(\boldsymbol{r}) - 1\right], \quad (27)$$

where $Q_{\alpha}(s,\boldsymbol{r})$ and $Q_{\alpha}{}^{\dagger}(s,\boldsymbol{r})$ are the path integrals for the $\alpha$-type polymer chain, $M_{\alpha}$ is the total number of $\alpha$-type polymer chains in the system, and $\phi_K(\boldsymbol{r})$ represents local volume fraction of $K$-type segments. The sum of the first two terms on the right-hand side of eq. (26) represents free energy contributions from the conformation entropy and the translational entropy of the chains. On the other hand, the last term $E$ represents the interaction energy between segments, where $\chi_{KK'}$ is Flory-Huggins interaction parameter between $K$-type and $K'$-type segments and $\gamma(\boldsymbol{r})$ is the Lagrange multiplier for the incompressibility condition ( $\sum_K \phi_K(\boldsymbol{r}) = 1$ ). Here, the path integrals $Q_\alpha(s, \boldsymbol{r})$ and ${Q_\alpha}^\dagger(s, \boldsymbol{r})$ can be obtained by solving similar equations as eqs.(24) and (25) with replacing the mean field $V(\boldsymbol{r})$ by $V_{K(s)}(\boldsymbol{r})$, where $K(s)$ is the segment species of the $s$-th segment.

In this study, our target is an A-B type semiflexible block copolymer melt where each polymer chain is composed of $N$ segments. We assume a symmetric diblock copolymer, where the first half of the chain $0 \leq s \leq N/2$ is the A sub-chain, and the other part $N/2 \leq s \leq N$ is the B sub-chain, respectively. In the framework of SCFT, the local volume fraction of $K$-type segments $\phi_K(\boldsymbol{r})$ and the mean field $V_K(\boldsymbol{r})$ are given as

$$\phi_K(\boldsymbol{r}) = M \int_{s \in \{s; K(s)=K\}} ds \, \frac{Q_\alpha(s, \boldsymbol{r}) {Q_\alpha}^\dagger(s, \boldsymbol{r})}{\int d\boldsymbol{r} \, Q_\alpha(s, \boldsymbol{r}) {Q_\alpha}^\dagger(s, \boldsymbol{r})}, \tag{28}$$

$$V_K(\boldsymbol{r}) = \sum_K \chi_{KK'} \phi_{K'}(\boldsymbol{r}) + \gamma(\boldsymbol{r}). \tag{29}$$

The path-integrals (eqs. (24) and (25)), the volume fraction (eq.(28)) and the mean field (eq.(29)) form a closed set of equations. To obtain the stable equilibrium state of phase separation under the constraint of fixed system size, iterative calculations for the coupled equations, eqs.(24), (25), (28) and (29), are required.

In our semiflexible Gaussian chain model of symmetric block copolymer chain, the model parameters are the total segment number per chain $N$, the Kuhn length $b$, the bending stiffness constant $k$, the Flory-Huggins interaction parameter $\chi_{\mathrm{AB}}$, and the temperature $k_\mathrm{B}T$. By using eqs. (20), (22) and (23), we can replace $b$ and $k$ by $\langle|\boldsymbol{u}|^2\rangle_0$ and $\langle\cos\theta\rangle_0^{\mathrm{eff}}$ and obtain the modified diffusion equations, eqs.(24) and (25). It should be noted here that both $\langle|\boldsymbol{u}|^2\rangle_0$ and $\langle\cos\theta\rangle_0^{\mathrm{eff}}$ depend on $k$. This means that changing the value of $k$ induces changes not only in the bond angle $\langle\cos\theta\rangle_0^{\mathrm{eff}}$ through the relation $\langle\cos\theta\rangle_0^{\mathrm{eff}} = \frac{kb^2}{3}$ (eq.(23)) but also in the average bond length $\langle|\boldsymbol{u}|^2\rangle_0$. To separate the effects of bond length and bond angle, we regard $\langle|\boldsymbol{u}|^2\rangle_0$ and $\langle\cos\theta\rangle_0^{\mathrm{eff}}$ as independent model parameters rather than to choose the usual set of parameters $b$ and $k$ as independent parameters. In such a framework, it will be convenient to fix the value of $\langle|\boldsymbol{u}|^2\rangle_0$ and non-dimensionalize eqs.(24) and (25) using the units of length $(\langle|\boldsymbol{u}|^2\rangle_0)^{1/2}$ and energy $k_\mathrm{B}T$ as

$$\frac{\partial\, Q_\alpha(s,\tilde{\boldsymbol{r}})}{\partial s} = \frac{1}{6}\left[\tilde{\nabla}^2 - \langle\cos\theta\rangle_0^{\mathrm{eff}} \frac{\tilde{\nabla} {Q_\alpha}^{\dagger}(s,\tilde{\boldsymbol{r}})}{Q_\alpha^{\dagger}(s,\tilde{\boldsymbol{r}})}\cdot\tilde{\nabla}\right] Q_\alpha(s,\tilde{\boldsymbol{r}}) - \tilde{V}_\alpha(\boldsymbol{r}) Q_\alpha(s,\tilde{\boldsymbol{r}}), \tag{30}$$

$$\frac{\partial\, Q_\alpha^{\dagger}(s,\tilde{\boldsymbol{r}})}{\partial s} = -\frac{1}{6}\left[\tilde{\nabla}^2 - \langle\cos\theta\rangle_0^{\mathrm{eff}} \frac{\tilde{\nabla} Q_\alpha(s,\tilde{\boldsymbol{r}})}{Q_\alpha(s,\tilde{\boldsymbol{r}})}\cdot\tilde{\nabla}\right] Q_\alpha^{\dagger}(s,\tilde{\boldsymbol{r}}) + \tilde{V}_\alpha(\tilde{\boldsymbol{r}})\, Q_\alpha^{\dagger}(s,\tilde{\boldsymbol{r}}), \tag{31}$$

where $\widetilde{\{*\}}$ represents non-dimensional form of variable $\{*\}$. In order to consider the effect of the bending stiffness on the phase separation, we change $\langle\cos\theta\rangle_0^{\mathrm{eff}}$ from -0.9 to 0.9, where $\langle\cos\theta\rangle_0^{\mathrm{eff}} = 0$ (or $k = 0$ from eq.(23)) corresponds to a flexible Gaussian chain without bending stiffness.

Using eqs. (30) and (31), we simulate the 1-dimensional lamellar structures formed by the symmetric semiflexible diblock polymer melts, where we assume that all segments have the same average bond angle $\langle\cos\theta\rangle_0^{\mathrm{eff}}$. We investigate the dependences of the lamellar period and the conformation per single polymer chain on this parameter $\langle\cos\theta\rangle_0^{\mathrm{eff}}$.

In our numerical simulations, spatial derivatives in eqs. (30) and (31) are evaluated using the finite difference method on a 1-dimensional mesh with a mesh width $\Delta x = 0.25$. The total chain length *N* is fixed at 200 which is discretized with a mesh width $\Delta s = 0.1$. The Flory-Huggins interaction parameter $\chi_{\mathrm{AB}}$ is fixed at 0.25 $(\chi_{\mathrm{AB}}N = 50)$ unless otherwise mentioned, while $\chi_{\mathrm{AA}} = \chi_{\mathrm{BB}} = 0$ in all calculations.

# III. Results and Discussions

Figure 2 shows the relationship between the lamellar period *D* and $\langle\cos\theta\rangle_0^{\mathrm{eff}}$ for an equilibrium lamellar phase where *D* on the vertical axis is normalized by the corresponding lamellar period for the flexible Gaussian chain model $\left(\langle\cos\theta\rangle_0^{\mathrm{eff}} =\right.$

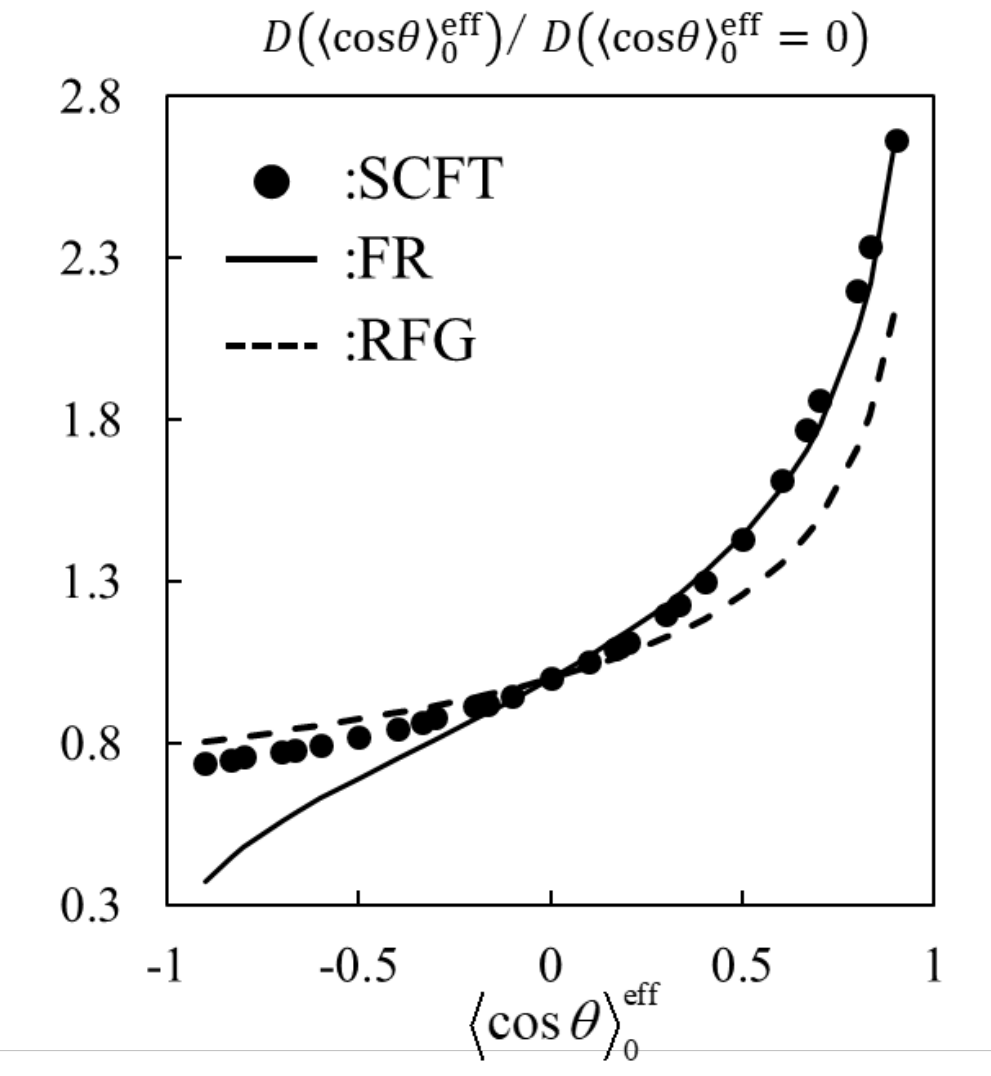

Figure 2. The relationship between the lamellar period *D* and $\langle\cos\theta\rangle_0^{\mathrm{eff}}$ in the range $-0.9 \leq \langle\cos\theta\rangle_0^{\mathrm{eff}} \leq 0.9$, where *D* in the vertical axis is normalized by the lamellar period of a flexible Gaussian chain model without bending stiffness $\left(\langle\cos\theta\rangle_0^{\mathrm{eff}} = 0 \text{ or } k = 0\right)$. The calculated results with our SCFT are represented by black dots. The solid and the dashed curves are theoretical predictions of the freely-rotating (FR) chain model and the renormalized flexible Gaussian chain (RFG) model.

0 or $k = 0$). The black dots in this figure are the results of our SCFT calculations for $-0.9 \leq \langle \cos\theta \rangle_0^{\mathrm{eff}} \leq 0.9$. These results represent that $D$ is an increasing function of $\langle \cos\theta \rangle_0^{\mathrm{eff}}$. When $\langle \cos\theta \rangle_0^{\mathrm{eff}}$ is positive ($k > 0$), $D$ changes significantly. On the other hand, when $\langle \cos\theta \rangle_0^{\mathrm{eff}}$ is negative ($k < 0$), $D$ is slightly varying from its value for the flexible Gaussian chain model. To understand such behavior of $D$ shown in Fig.2, we here give an analytic expression of the relation between $D$ and $\langle \cos\theta \rangle_0^{\mathrm{eff}}$. Our analytic expression is based on the strong segregation theory of microphase separation of block copolymer melt, where the entropy of the chain stretching is estimated using the flexible Gaussian chain model, or the so-called freely rotating (FR) chain model where both the bond length and the bond angle are fixed.

According to the strong segregation theory, we obtain the relationship $D \propto \rho_0^{-1/3}(bN)^{2/3}$, where $\rho_0$ is the total number density of the segments [54]. When we fit this expression of $D$ to our SCFT data, it is important to use the Kuhn statistical length $b$ which is renormalized by the effects of the bending stiffness energy (i.e. the 2nd term in the right-hand side of eq.(5) as was discussed in the previous section). Within the framework of the flexible Gaussian chain model, this renormalization is expressed by the first term on the right-hand side of eq. (14) in which this first term and the second term represent the coarse-grained bond vector and curvature vector made of two adjacent bond vectors, respectively. From the coefficient of the first term of eq. (14), we can evaluate the effective Kuhn statical length $b_{\mathrm{RFG}}$ using the relation of $\frac{3k_{\mathrm{B}}T}{2b_{\mathrm{RFG}}^2} = \frac{3k_{\mathrm{B}}T}{2}\left(\frac{1}{b_0^2} - \frac{k_0}{3}\right)$, where the subscript RFG stands for the "renormalized flexible Gaussian chain model". Such an approximation leads to the expression of the effective Kuhn statistical length $b_{\mathrm{RFG}} = b/(1 - kb^2/(3k_{\mathrm{B}}T))^{1/2} = b/\left(1 - \langle\cos\theta\rangle_0^{\mathrm{eff}}\right)^{1/2}$, where we used eq. (23). In this "RFG" model, the internal energy of a single chain is described by the flexible Gaussian chain with renormalized bond energy.

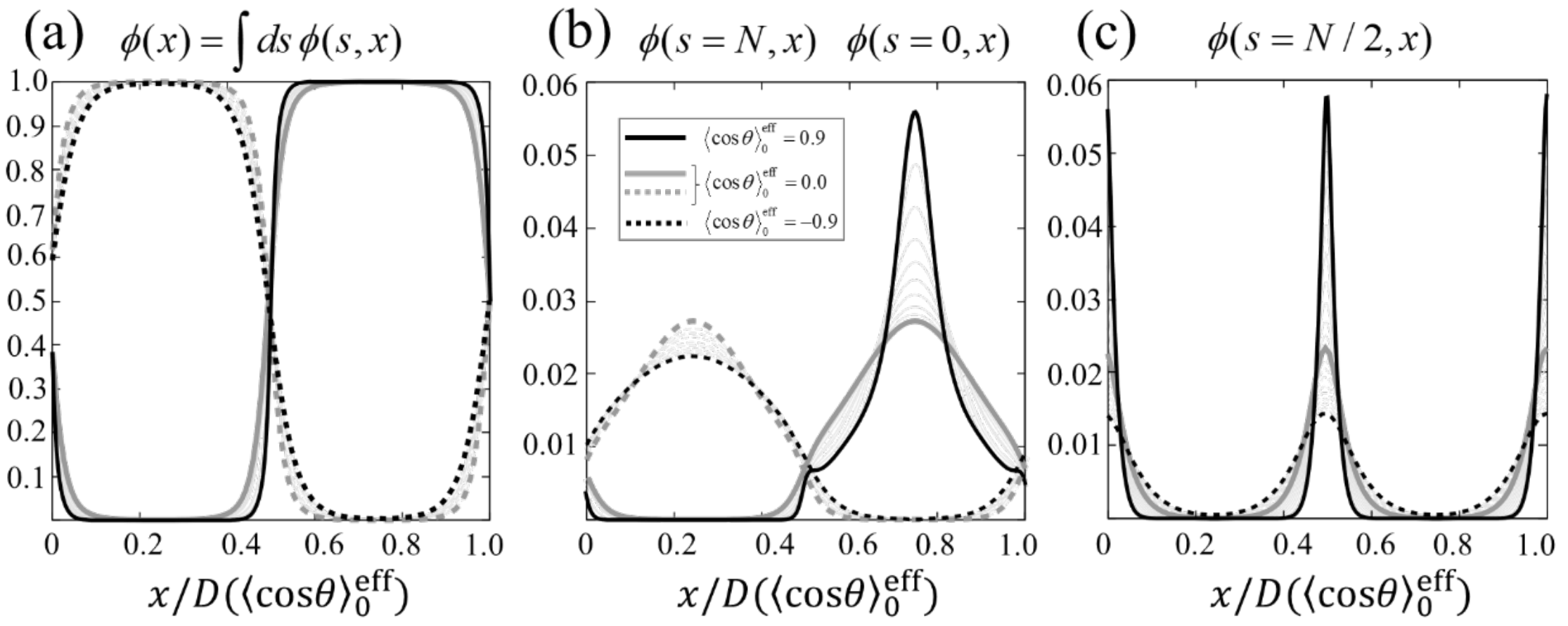

Figure 3. Profiles of the segment density distribution in equilibrium state for various values of $\langle\cos\theta\rangle_0^{\mathrm{eff}}$. (a) Total segment density of A and B, $\phi(x)$, (b) end segment densities $\phi(s = N, x)$ and $\phi(s = 0, x)$, and (c) center segment density $\phi(s = N/2, x)$. These profiles are plotted for $\langle\cos\theta\rangle_0^{\mathrm{eff}}$ from $-0.9$ to 0.9 with intervals of 0.1. For $\langle\cos\theta\rangle_0^{\mathrm{eff}} > 0$, the density profiles of the A-type segment are shown with solid lines, while the densities of the B-type segment are shown with dashed lines for $\langle\cos\theta\rangle_0^{\mathrm{eff}} < 0$. The profiles for $\langle\cos\theta\rangle_0^{\mathrm{eff}} = -0.9, 0.0$, and 0.9 are highlighted with bold lines colored with black for $\langle\cos\theta\rangle_0^{\mathrm{eff}} = -0.9$, dark gray for $\langle\cos\theta\rangle_0^{\mathrm{eff}} = 0.0$, and black for $\langle\cos\theta\rangle_0^{\mathrm{eff}} = 0.9$, respectively. All other profiles are colored with light gray. The horizontal axis is normalized using $D\left(\langle\cos\theta\rangle_0^{\mathrm{eff}}\right)$ so that its range is between 0 to 1.

On the other hand, we consider the modified Kuhn length based on the FR chain model, in which adjacent segments with bond length $b$ are jointed by an angle $\theta$ and can rotate freely keeping $\theta$ constant. Since the statistical average of the end to end distance $R$ of the FR chain is expressed by $R \approx N^{1/2}b((1-\cos\theta)/(1+\cos\theta))^{1/2}$ [55], the modified Kuhn statistical length for the FR chain can be estimated as $b_{\mathrm{FR}} = b((1+\cos\theta)/(1-\cos\theta))^{1/2} \sim b\left(\left(1+\langle\cos\theta\rangle_0^{\mathrm{eff}}\right)/\left(1-\langle\cos\theta\rangle_0^{\mathrm{eff}}\right)\right)^{1/2}$, where $\cos\theta$ was replaced by $\langle\cos\theta\rangle_0^{\mathrm{eff}}$. By combining the estimation of the lamellar period ($D \propto \rho_0^{-1/3}(bN)^{2/3}$) obtained with the strong segregation theory and the expression of the modified Kuhn length ($b_{\mathrm{RFG}}$ or $b_{\mathrm{FR}}$), an analytic expression for the relationship between $D$ and $\langle\cos\theta\rangle_0^{\mathrm{eff}}$ is obtained.

Based on the above discussion, we show in Fig. 2 the analytic relation between $D$ and $\langle\cos\theta\rangle_0^{\mathrm{eff}}$ for RFG chain model (dashed line) and FR chain model (solid line). From this figure, we recognize that our SCFT results (black dots) can be quantitatively fitted by the FR model (the solid line) for $\langle\cos\theta\rangle_0^{\mathrm{eff}} \geq 0$ and the RFG chain model (the dashed line) for $\langle\cos\theta\rangle_0^{\mathrm{eff}} \leq 0$, which indicates a crossover at $\langle\cos\theta\rangle_0^{\mathrm{eff}} = 0$ in the single polymer chain conformation between those two regimes represented by the RFG chain model and the FR chain model. This crossover can be understood qualitatively as follows. For $\langle\cos\theta\rangle_0^{\mathrm{eff}} > 0$ ($k > 0$), the chain is semiflexible and has a positive persistence length which is larger than the vanishing persistence length of the flexible Gaussian chain model. As a result, the chain is stretched, which leads to the larger lamellar domain spacing than the flexible Gaussian chain model. This behavior is just the same as the FR chain model. On the other hand, when $\langle \cos\theta \rangle_0^{\mathrm{eff}} < 0$ $(k < 0)$, the bending stiffness energy produces a negative correlation between the directions of adjacent bonds, which corresponds to a negative persistence length. However, such a negative persistence length is meaningless, and the chain has flexible Gaussian chain nature that has vanishing persistence length. Therefore, the lamellar domain spacing is unchanged from its value for the flexible Gaussian chain model.

Next, we investigate the conformation of a single polymer chain in the equilibrium lamella structure. Figure 3 shows the equilibrium local volume fractions of the segments for various values of $\langle \cos\theta \rangle_0^{\mathrm{eff}}$. The horizontal axis represents the coordinate perpendicular to the lamella interface which is normalized by the lamellar period $D\left(\langle \cos\theta \rangle_0^{\mathrm{eff}}\right)$ so that the minimum and the maximum values are 0 and 1, respectively. In Fig. 3 and the following figures, to save the space, we utilize the symmetry between A and B by showing only halves of the segment distributions $\phi_{\mathrm{A}}$ and $\phi_{\mathrm{B}}$, i.e. we show $\phi_{\mathrm{A}}$ for $\langle \cos\theta \rangle_0^{\mathrm{eff}} > 0$ and $\phi_{\mathrm{B}}$ for $\langle \cos\theta \rangle_0^{\mathrm{eff}} < 0$, respectively. Then, the profiles of $\phi_{\mathrm{A}}$ and $\phi_{\mathrm{B}}$ for $\langle \cos\theta \rangle_0^{\mathrm{eff}} < 0$ and $\langle \cos\theta \rangle_0^{\mathrm{eff}} > 0$ can be obtained by the symmetry $\phi_{\mathrm{A}}(x) = 1 - \phi_{\mathrm{B}}(x)$. Figure 3 (a) represents density profiles of $\phi_{\mathrm{A}}(x) = \int_0^{N/2} \phi_{\mathrm{A}}(s,x)ds$ and $\phi_{\mathrm{B}}(x) = \int_{N/2}^{N} \phi_{\mathrm{B}}(s,x)ds$. This figure shows that two interfaces of the phase separation are observed at $\frac{x}{D\left(\langle \cos\theta \rangle_0^{\mathrm{eff}}\right)} = 0.5$ and 0 (or 1.0), where $\phi_{\mathrm{A}}$ distributes in the region of $x/D\left(\langle \cos\theta \rangle_0^{\mathrm{eff}}\right) > 0.5$, and $\phi_{\mathrm{B}}$ in $x/D\left(\langle \cos\theta \rangle_0^{\mathrm{eff}}\right) < 0.5$, respectively. Figure 3 (b) represents density profiles of end segments $\phi_{\rm A}(s=0,x)$ and $\phi_{\rm B}(s=N,x)$. The end segments are populated at the center of each domain, i.e., $x/D(\langle\cos\theta\rangle_0^{\rm eff})=0.25$ for $\phi_{\rm B}(s=N,x)$ and $x/D(\langle\cos\theta\rangle_0^{\rm eff})=0.75$ for $\phi_{\rm A}(s=0,x)$. Figure 3 (c) shows density profiles of central segments, i.e., $\phi_{\rm A}(s=N/2,x)=\phi_{\rm B}(s=N/2,x)$. The central segments are localized at the lamella interfaces $\left(\frac{x}{D(\langle\cos\theta\rangle_0^{\rm eff})}=0.0, 0.5 \text{ and } 1.0\right)$. The dependences of these data on $\langle\cos\theta\rangle_0^{\rm eff}$ clearly indicate that the polymer conformations in the equilibrium state are influenced by the polymer stiffness $\langle\cos\theta\rangle_0^{\rm eff}$. When $\langle\cos\theta\rangle_0^{\rm eff}$ is positive and its magnitude increases, the peaks in the segment density distributions become sharper, which means the effects of thermal fluctuation are decreased. This is due to the less conformation entropy of stiff polymer chains compared with the flexible Gaussian chains because the number of statistically independent segments becomes fewer as the persistence length increases. On the other hand, when $\langle\cos\theta\rangle_0^{\rm eff}$ is negative and its magnitude increases, the distributions of each segment become slightly broader but essentially unchanged. This is because the behavior of the chain for $\langle\cos\theta\rangle_0^{\rm eff}<0$ is the same as that for $\langle\cos\theta\rangle_0^{\rm eff}=0$ as discussed concerning Fig.2. For the better understanding of the dependences of the polymer conformations on $\langle\cos\theta\rangle_0^{\rm eff}$, we will discuss the local segment orientations in the following.

The probability distribution of bond vectors is described by the vector-order parameter defined by

$$\boldsymbol{\psi}(s, \boldsymbol{x}) = \langle \delta(\boldsymbol{x} - \boldsymbol{r}(s))\boldsymbol{u}(s) \rangle, \tag{32}$$

where $\langle * \rangle$ represents the statistical average and $\boldsymbol{r}(s)$ and $\boldsymbol{u}(s)$ represent the segment position and the bond vector at $s$. Using the path-integrals, the vector-order parameter is expressed as

$$\boldsymbol{\psi}(s, \boldsymbol{x}) = \frac{M}{Z(1)}\left[Q(s, \boldsymbol{x})\nabla Q^{\dagger}(N - s, \boldsymbol{x}) - \nabla Q(s, \boldsymbol{x})Q^{\dagger}(N - s, \boldsymbol{x})\right]. \tag{33}$$

The derivation of this equation is shown in Ref. [56]. Here note that $\boldsymbol{\psi}(s, \boldsymbol{x})$ defined by eq. (33) can take either positive or negative value depending on the bond orientation. In our simulation, $\boldsymbol{\psi}(s)$ is positive (negative) when the bond vector from the segment at $s$ to the segment at ($s$ + d$s$) is in the positive (negative) direction of the $x$-axis. Figure 4 represents the vector-order parameters $\boldsymbol{\psi}(s,x)$, which are obtained with the same conditions as those in Fig. 3. Figure 4 (a) represents the sum of all bond orientations for all A-type segments (solid line) and B-type segments (dashed line). Here, we again utilize the symmetry between A and B species to reduce the space, i.e., $\int_0^{N/2} ds\, \boldsymbol{\psi}(s,x)$ of A-type segments are shown for $\langle \cos\theta \rangle_0^{\text{eff}} > 0$, while $\int_{N/2}^{N} ds\, \boldsymbol{\psi}(s,x)$ of B-type segments are shown for $\langle \cos\theta \rangle_0^{\text{eff}} < 0$, respectively. For $\langle \cos\theta \rangle_0^{\text{eff}} > 0$, the amplitude of $\boldsymbol{\psi}(s,x)$ increases with $\langle \cos\theta \rangle_0^{\text{eff}}$, which represents the increase of the population of the polymer segments aligning in the perpendicular direction to the lamella interfaces. Our data on the profile of $\boldsymbol{\psi}(s,x)$ is consistent with those reported in the previous research, where the sharp peaks are found around the interface and the profiles decrease rapidly toward 0 around the center of each domain [57]. For $\langle \cos\theta \rangle_0^{\text{eff}} < 0$, the same tendency in the bond orientation as for the case with $\langle \cos\theta \rangle_0^{\text{eff}} > 0$ is found as shown in the inset of Fig. 4 (a).

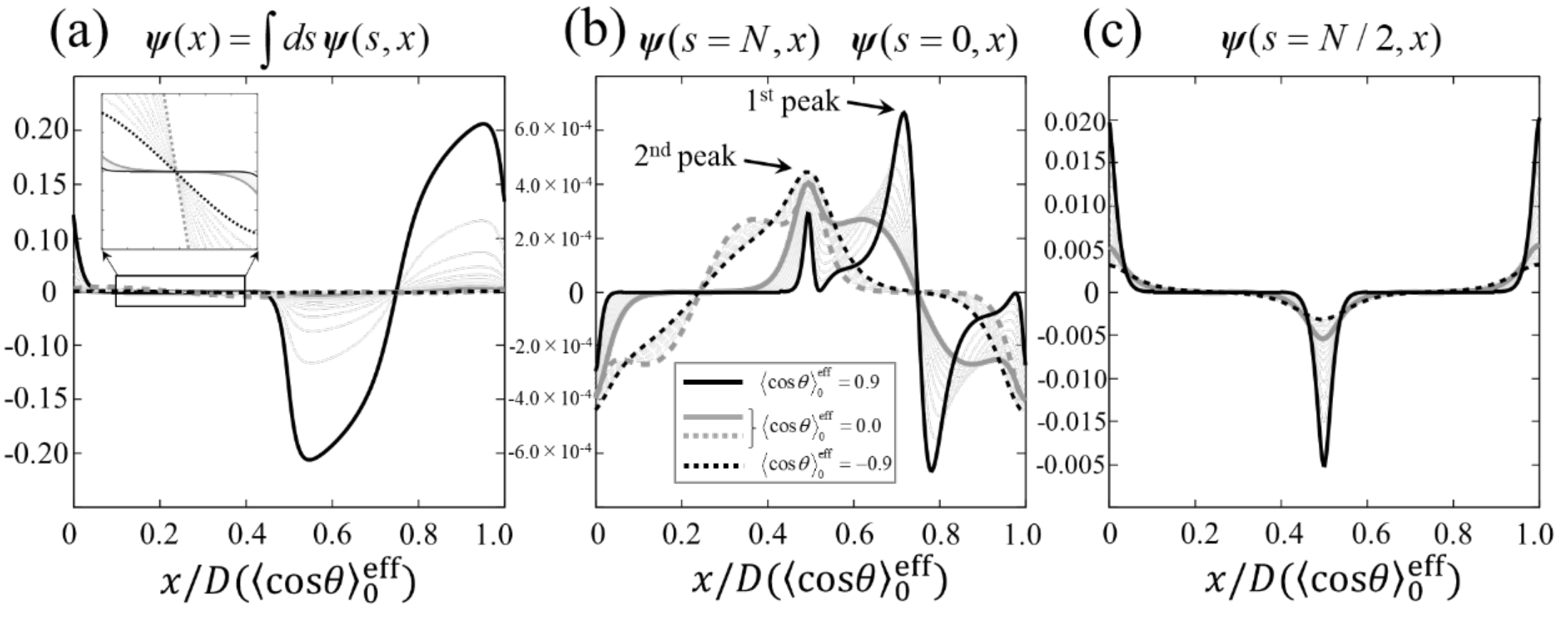

Figure 4. Profiles of the segment orientation distributions in the equilibrium state for various values of $\langle\cos\theta\rangle_0^{\mathrm{eff}}$: (a) sum of $\boldsymbol{\psi}(x)$ for all segments, (b) $\boldsymbol{\psi}(x)$ for the ends segments $\boldsymbol{\psi}(s = N, x)$ and $\boldsymbol{\psi}(s = 0, x)$, and (c) $\boldsymbol{\psi}(x)$ for the center segment $\boldsymbol{\psi}(s = N/2, x)$. Inset in figure (a) is an enlarged illustration for the part surrounded by a square. These profiles are plotted from $\langle\cos\theta\rangle_0^{\mathrm{eff}} = -0.9$ to $\langle\cos\theta\rangle_0^{\mathrm{eff}} = 0.9$ with intervals of 0.1. The profiles for $\langle\cos\theta\rangle_0^{\mathrm{eff}} = -0.9, 0.0, 0.9$ are highlighted with bold lines colored with black for $\langle\cos\theta\rangle_0^{\mathrm{eff}} = -0.9$, dark gray for $\langle\cos\theta\rangle_0^{\mathrm{eff}} = 0.0$, and black for $\langle\cos\theta\rangle_0^{\mathrm{eff}} = 0.9$. For $\langle\cos\theta\rangle_0^{\mathrm{eff}} > 0$, the orientation of the A-type segment is shown with a solid line, while a dashed line is used for B-type segment for $\langle\cos\theta\rangle_0^{\mathrm{eff}} < 0$. The system size is normalized by the lamellar domain spacing $D(\langle\cos\theta\rangle_0^{\mathrm{eff}})$ so that the horizontal axis ranges from 0 to 1.

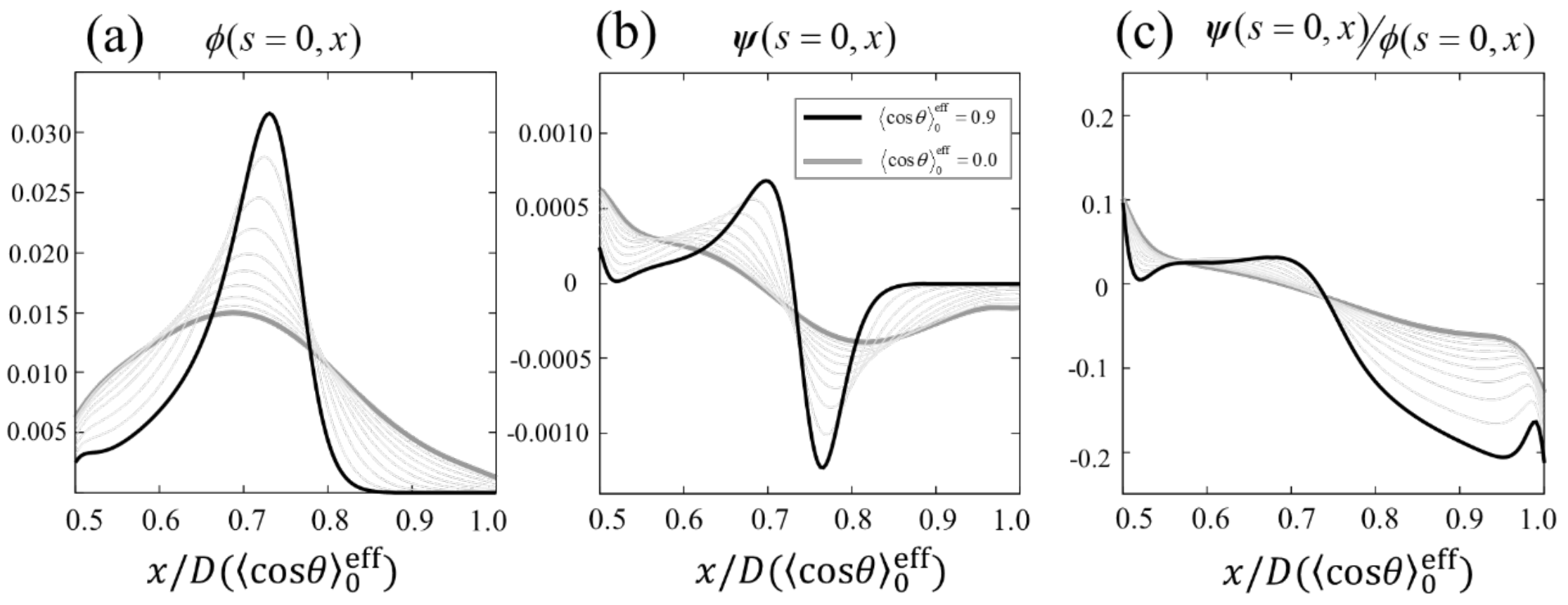

Figure 5. The order parameters of the end segment (*s*=0) when the center segment (*s*=*N*/2) is fixed at $x/D\left(\langle\cos\theta\rangle_0^{\text{eff}}\right)=0.5$ (phase separation interface): (a) the segment density $\phi(s=0,x)$, (b) the vector order parameter $\boldsymbol{\psi}(s=0,x)$ and (c) normalized vector order parameter $\boldsymbol{\psi}(s=0,x)/\phi(s=0,x)$. These profiles are plotted for $\langle\cos\theta\rangle_0^{\text{eff}}$ from $\langle\cos\theta\rangle_0^{\text{eff}}=0.0$ to $\langle\cos\theta\rangle_0^{\text{eff}}=0.9$ with intervals of 0.1. The profiles for $\langle\cos\theta\rangle_0^{\text{eff}}=0.0$, and 0.9 are highlighted with bold lines colored with dark gray for $\langle\cos\theta\rangle_0^{\text{eff}}=0.0$, black for $\langle\cos\theta\rangle_0^{\text{eff}}=0.9$, respectively. All other profiles are colored with light gray. The horizontal axis covers the range $0.5<x/D\left(\langle\cos\theta\rangle_0^{\text{eff}}\right)<1.0$.

However, the amplitude is much smaller than that for $\langle\cos\theta\rangle_0^{\text{eff}}>0$, which means that the bond distribution of the polymer segments for $\langle\cos\theta\rangle_0^{\text{eff}}<0$ is almost isotropic.

Figures 4 (b) and (c) represent the vector order parameters $\boldsymbol{\psi}(s,x)$ for the segments at the chain ends and at the chain center, respectively. Near the interfaces of the phase-separated domains, the end-segment has positive peak in $\boldsymbol{\psi}$ (denoted as the 1st peak in Fig.4(b)), and the central segment has the negative peak in $\boldsymbol{\psi}$ (Fig.4(c)). The averaged bond-vector shown in Fig. 4 (a) has the same direction as that for the central segment. It should also be noted that the amplitude of the peak of the central segment is much larger than those of the end-segments. Therefore, the orientation of the bond-vector around the central segment greatly contributes to the average bond vector in comparison to the bond vectors near the end-segments. As a conclusion from Figs. 3 (c) and 4 (c), the central segment is concentrated around the interface, and its bond is aligned in the perpendicular direction to the interface. These concentrations are strengthened as $\langle \cos\theta \rangle_0^{\mathrm{eff}}$ increases.

For the bonds at the end segments, their directions behave in a more complex manner than that of the central segment. Figure 4 (b) shows that $\boldsymbol{\psi}$ of the end-segment has two independent peaks near the interface ($1^{\mathrm{st}}$-peak) and near the center of each domain ($2^{\mathrm{nd}}$-peak), while the center segment has only one peak near the interface. The $1^{\mathrm{st}}$ peak decreases as $\langle \cos\theta \rangle_0^{\mathrm{eff}}$ increases. On the other hand, in the case of the $2^{\mathrm{nd}}$ peak, its amplitude increases with $\langle \cos\theta \rangle_0^{\mathrm{eff}}$ while shifting the peak position from the interface to the center of each domain. To investigate the conformations of the end-segments related to this complex behavior of $\boldsymbol{\psi}$, we show in Fig. 5 the order parameters of the end-segment in the case that the center segment ($s=N/2$) is fixed at the interface region ($x/D\left(\langle \cos\theta \rangle_0^{\mathrm{eff}}\right) = 0.5$). As shown in Fig. 5 (a), the end-segment for small bending stiffness is broadly distributed, however, that for large bending stiffness is localized around the bulk region of the domain. The amount of the end segment at the interface gradually decreases with the increase of the bending stiffness, which corresponds to the behavior of the $1^{\mathrm{st}}$-peak around the interface shown in Fig. 4 (b). The peak at the middle of the domain gradually grows while shifting its position to the center part of the domain, which implies that the polymer chain conformation is extended toward the center region of the domain. Figure 5 (b) shows the positive and the negative peaks of $\boldsymbol{\psi}(s = 0, x)$ at $x/D(\langle\cos\theta\rangle_0^{\mathrm{eff}}) = 0.7$ and 0.75, respectively. These peaks behave similarly to those of 2nd peaks in Fig. 4(b), representing not only polymers with folded conformations but also those with extended conformations as explained in the following. This negative peak indicates extension of the polymer chains because the negative $\boldsymbol{\psi}(s = 0, x)$ means that the end-segments are oriented toward the positive direction of the horizontal axis on average. The degree of the extension increases with the increase in the bending stiffness because $\boldsymbol{\psi}(s = 0, x)/\phi(s = 0, x)$, i.e. the net direction of the bond normalized by the amount of the total segments, also increases with the increase in the bending stiffness as shown in Fig. 5 (c). Therefore, the bending stiffness affects the statistical distribution of the polymer conformations so as to increase the population of the polymer chains that extend toward the perpendicular direction to the lamellar interface.

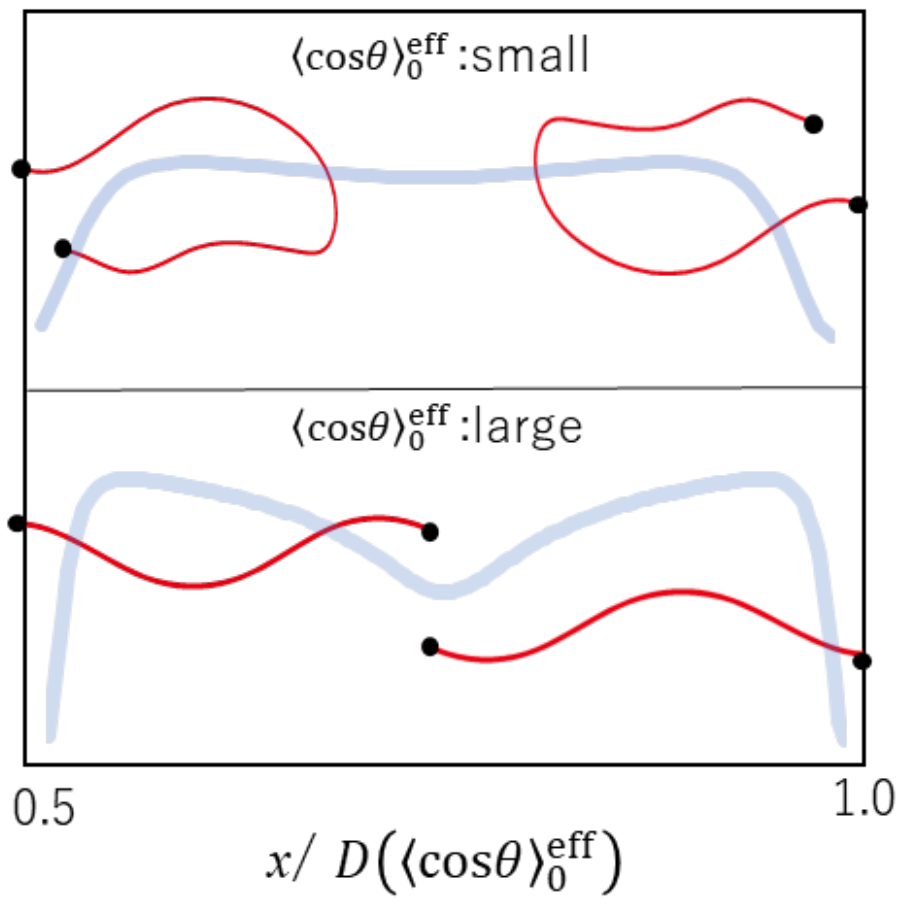

Figure 6. Schematic pictures of polymer conformation (red curves) and the calculated results of the profile of the chemical potential $\gamma(x)$ (blue curves) for large (top) and small (bottoms) $\langle\cos\theta\rangle_0^{\mathrm{eff}}$ cases. The horizontal axis represents normalized coordinates. The phase separation interfaces are placed at 0.5 and 1.0 of $x/D(\langle\cos\theta\rangle_0^{\mathrm{eff}})$. The edge segment and central segment are denoted by a black dot with segment number 0 and $N/2$, respectively. The blue curves represent the distribution of the chemical potential with the case of $\langle\cos\theta\rangle_0^{\mathrm{eff}} = 0.0$ (top) and 0.9 (bottom), respectively.

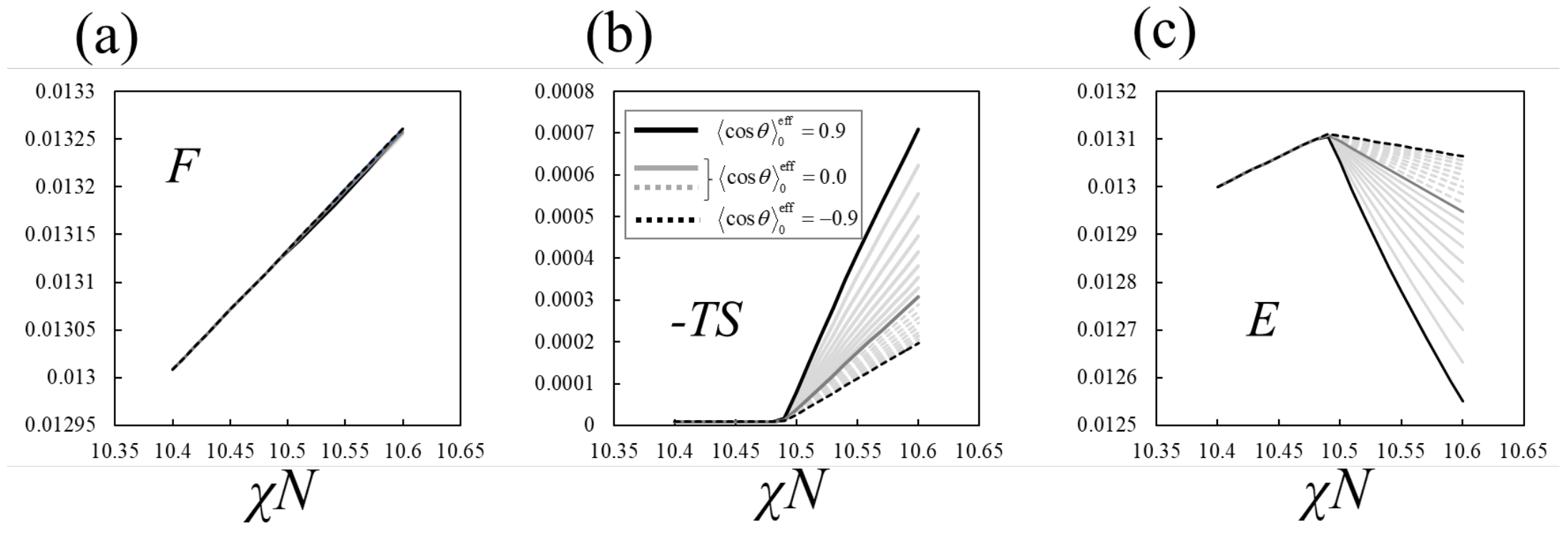

Figure 7. Free energy components for various values of $\langle\cos\theta\rangle_0^{\text{eff}}$ for $\chi N$ around the order-disorder transition. (a) total free energy $F$, (b) entropy -$TS$, and (c) internal energy $E$, obtained with eqs. (27) and (28) given in section II. For $\langle\cos\theta\rangle_0^{\text{eff}} > 0$, free energy profiles are shown with solid lines, while the free energies of the B-type segment are shown with dashed lines for $\langle\cos\theta\rangle_0^{\text{eff}} < 0$. The profiles for $\langle\cos\theta\rangle_0^{\text{eff}} = -0.9, 0.0$, and 0.9 are highlighted with bold lines colored with black for $\langle\cos\theta\rangle_0^{\text{eff}} = -0.9$, dark gray for $\langle\cos\theta\rangle_0^{\text{eff}} = 0.0$, and black for $\langle\cos\theta\rangle_0^{\text{eff}} = 0.9$, respectively. All other profiles are colored with light gray.

Schematic pictures of the polymer conformations for smaller and larger values of the bending stiffness $\langle\cos\theta\rangle_0^{\text{eff}}$ are shown in Fig. 6. In these figures, the SCFT simulation results of the chemical potential $\gamma(x)$ are also shown by thick curves. In both cases there are potential minima in $\gamma(x)$ at the interface and at the middle of the domain where the end-segments gather because the end segments are more influenced by the potential field than the segments that are connected to two bonds on its both sides. In the case of the smaller bending stiffness, potential well at the middle of the lamellar domain is shallow compared to the case with the larger bending stiffness. Thus, the polymer chains are folded because the end segments are attracted to the interfacial region by the forces due to the potential minimum at the interface. On the other hand, in the case of the larger bending stiffness, polymers extend toward the center region of the lamellar domain because the segments are attracted by the force originated from the deep potential well at the center of the lamellar domain.

So far, we studied the phase separation in the strong segregation regime ($\chi N = 50$). Next, we consider the behavior of the free energy in the week segregation regime, especially around the order-disorder transition point. Figure 7 shows the components of the free energy for various values of $\langle \cos\theta \rangle_0^{\mathrm{eff}}$, where the value of $\chi N$ changes by changing $\chi$ for constant chain length *N* =200. From Fig. 7 (a), the total free energy (*F*) smoothly increases with $\chi N$. On the other hand, the entropy component (-*TS*, Fig. 7 (b)) and the internal energy component ($E$, Fig. 7 (c)) show a transition between two linear dependences, which indicates that a second-order phase transition occurs at the point where the derivative coefficient is discontinuous. For all $\langle \cos\theta \rangle_0^{\mathrm{eff}}$, the transition points are the same at $\chi N = 10.49$, which coincides with the critical point for the standard flexible Gaussian chain model. Here, we validate this critical point does not depend on $\langle \cos\theta \rangle_0^{\mathrm{eff}}$, in other words, our introducing bending stiffness of the single polymer chain

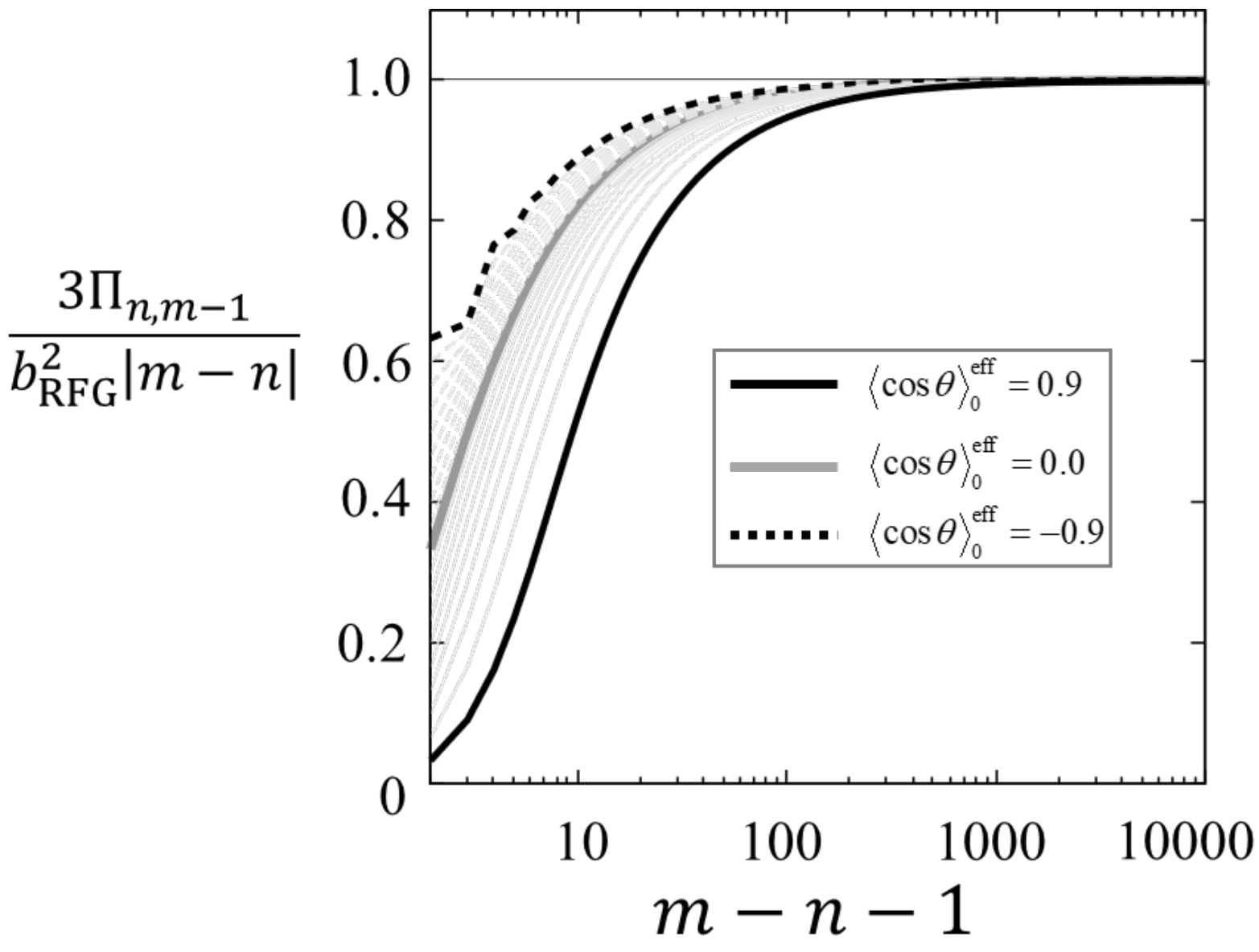

Figure 8. The normalized variance of the two-point correlation function for difference of $\langle\cos\theta\rangle_0^{\mathrm{eff}}$s. For $\langle\cos\theta\rangle_0^{\mathrm{eff}} > 0$, free energy profiles are shown with solid lines, while the free energies of the B-type segment are shown with dashed lines for $\langle\cos\theta\rangle_0^{\mathrm{eff}} < 0$. The profiles for $\langle\cos\theta\rangle_0^{\mathrm{eff}} = -0.9, 0.0$, and 0.9 are highlighted with bold lines colored with black for $\langle\cos\theta\rangle_0^{\mathrm{eff}} = -0.9$, dark gray for $\langle\cos\theta\rangle_0^{\mathrm{eff}} = 0.0$, and black for $\langle\cos\theta\rangle_0^{\mathrm{eff}} = 0.9$, respectively. All other profiles are colored with light gray.

does not shift the critical point from that of the flexible Gaussian chain model.

As is derived in Appendix B, based on the Hamiltonian of the semiflexible Gaussian chain model without any external potential, the probability $P(n,\boldsymbol{r};m,\boldsymbol{r}')$ of finding *n*-th and *m*-th segments at ***r*** and ***r'*** is obtained by

$$P(n,\boldsymbol{r};m,\boldsymbol{r}') = const.\times \exp\left[-\frac{|\boldsymbol{r}'-\boldsymbol{r}|^2}{2\Pi_{n,m-1}}\right], \tag{34}$$

$$\Pi_{n,m-1} = \frac{1}{\beta(m-n-1)}\left[\sum_{l=1}^{m-n-2}\frac{1}{\tilde{a}+2\tilde{b}\cos\frac{\pi}{m-n-1}l}\left[\frac{1-(-1)^l}{2}\frac{1+\cos\left\{\frac{\pi}{m-n-1}l\right\}}{\sin\left\{\frac{\pi}{m-n-1}l\right\}}\right]^2\right], \tag{35}$$

where $\tilde{a} = 3k_{\mathrm{B}}T/2b_0^2$ and $\tilde{b} = -k_0/4$. By coupling eqs.(35) and (23), the dependence of $\Pi_{n,m-1}$ on difference between *m*-th and *n*-th segments are plotted for various values of $\langle\cos\theta\rangle_0^{\text{eff}}$ in Figure 8. The vertical axis of this figure represents $\Pi_{n,m-1}$ normalized with $b_{\text{RFG}}^2|m-n|/3$, where $b_{\text{RFG}}$ is effective Kohn statistical length for RFG model. Since our framework of SCFT is continuous model of contour segment $s$ by taking the limit $ds \rightarrow 0$, which corresponds to the case of $m-n-1 \rightarrow \infty$ in Figure 8. All plots with different $\langle\cos\theta\rangle_0^{\text{eff}}$ converge to the same value of 1.0 if *m-n* is large enough. Therefore, the Fourier transform of the two-point correlation function of the semiflexible Gaussian chain model is regarded as the same Debye function as that of the flexible Gaussian chain model, with a difference only in the Kuhn length. Therefore, based on the random phase approximation, the semiflexible Gaussian chain model has the same critical point in $\chi N$ as that of the flexible Gaussian chain model.

# IV. Conclusion

We investigated the phase separation phenomena of a semiflexible block-copolymer melt using the SCFT coupled with semiflexible Gaussian chain model. We proposed a Hamiltonian for the semiflexible polymer chains including not only the harmonic spring energy of bonds but also directional correlations between adjacent bonds (*i.e.*, the bending stiffness). Based on this Hamiltonian, we derived modified diffusion equations for the path-integrals in forward and backward directions along the polymer chain. In these modified diffusion equations, the forward and the backward path-integrals are coupled with each other, which is different from the situations in the flexible bead-spring model or the worm-like chain model.

Using our model for the SCFT, we investigated symmetric semiflexible di-block copolymers, where we found that the bending stiffness strongly affects the polymer conformations and the phase-separated structures, especially in the strong segregation region. The period of the equilibrium lamellar domains increases with the increase in the bending stiffness. The results of the SCFT on the lamella period for various values of bending stiffness are fitted by the theoretically predicted curves based on the freely-rotating (FR) chain model and renormalized flexible Gaussian (RFG) chain model. We also found that these curves show a crossover at vanishing bending stiffness. By the analysis of the polymer conformations, folded polymer chains for negative bending stiffness start to extend in a direction perpendicular to the lamella interface when the bending stiffness becomes positive. This bending stiffness does not affect the order-disorder transition point.

Here we note that there are some computational models which consider the bending stiffness of polymers for the phase separated structures. Grason et al. introduced Frank's elastic energy into polymer chains based on the SCFT and simulated the twisted domains in phase separation of block copolymer system [58]. In their studies, the elastic energy is expressed as the intermolecular interactions and the individual polymer chains are modeled as the standard Gaussian chain, which differs from the polymer chain model proposed in this study. Another computational model is particle-field hybrid simulation, proposed by Milano and Kawakatsu, in which the intramolecular and intermolecular interactions are represented in the same manner as molecular dynamics simulation (MD) and SCFT, respectively [59]. Their model called MD-SCFT hybrid model enables us to reproduce detailed molecular structure by potential energies associated with the spring, bending and dihedral interactions of adjacent 2, 3 and 4 atoms, while decreasing the computational cost compared with standard MD simulation with the help of the mean field for intermolecular interactions. Compared with their model, our proposed model has the advantage of being able to quantitatively evaluate the stability of phase-separated structures based on the free energy evaluations. Furthermore, the dihedral angular potential used in MD-SCFT hybrid model, which is omitted in the semiflexible Gaussian chain model, can be considered for the SCFT calculation by coupling of forward and backward path-integrals of the standard worm-like chain model, which will be reported in our forthcoming paper.

## Appendix. A

This section discusses the domain of $\langle\cos\theta\rangle_0 = kb^2/3$ (eq. (23)), which is given as $-1 < \langle\cos\theta\rangle_0 < 1 \left(-\frac{3}{b^2} < k < \frac{3}{b^2}\right)$. As described in the main text, $g_{\pm}$(eq. (22)) diverges when $k = \frac{3}{b^2}$ or $-\frac{3}{b^2}$, and the Hamiltonian of the polymer chain becomes equivalent to the worm-like chain model. Therefore, to keep the chain statistic within the Gaussian chain regime, the bending elasticity constant $k$ must take values within the domain $-\frac{3}{b^2} < k < \frac{3}{b^2}$. However, the bending elasticity constant ($\kappa$) found in the worm-like chain model can theoretically take any positive real number. The purpose of this section is to elucidate the relationship between the bending elasticity constant $k$ in the semiflexible Gaussian chain model and $\kappa$. Specifically, it will be demonstrated that $-1 < \langle\cos\theta\rangle_0 < 1 \left(-\frac{3}{b^2} < k < \frac{3}{b^2}\right)$ is satisfied when $-\infty < \kappa < \infty$.

In the worm-like chain model, the bending elastic energy between three consecutive segments centered at the *i*-th segment is expressed as the product of adjacent unit bond vectors, *i.e.*, $\kappa\,\hat{\boldsymbol{u}}_i \cdot \hat{\boldsymbol{u}}_{i+1}$. However, in the semiflexible Gaussian chain model, the length of the bond vectors cannot be fixed, and the bending elastic energy is replaced to $\kappa\,\boldsymbol{u}_i \cdot \boldsymbol{u}_{i+1}/\langle|\boldsymbol{u}|^2\rangle_0$. Consequently, we obtain the relationship between the bending elastic constants for semiflexible Gaussian chain and worm-like chain models as $k = \kappa/\langle|\boldsymbol{u}|^2\rangle_0$. We substitute this relation into the following equation as

$$\langle|\boldsymbol{u}|^2\rangle_0 = \frac{3}{2}\left(\frac{b^2}{3-kb^2} + \frac{b^2}{3+kb^2}\right), \quad \text{(A-1)}$$

which is obtained by eqs.(20) and (22). We can express $\langle|\boldsymbol{u}|^2\rangle_0$ as a function of $\kappa$ as follows,

$$\langle|\boldsymbol{u}|^2\rangle_0 = \frac{1}{2}b^2 + \frac{1}{2}\left(1 + \frac{4}{9}\kappa^2\right)^{\frac{1}{2}} b^2. \quad \text{(A-2)}$$

Equation (A-2) indicates that the statistical average of the square of a bond vector is linearly dependent on $\kappa^2$, when $\kappa$ is sufficiently small. By substituting eq. (A-1) into $k = \kappa/\langle|\boldsymbol{u}|^2\rangle_0$, we obtain the following relation.

$$k = \frac{\kappa}{\frac{1}{2}b^2 + \frac{1}{2}\left(1 + \frac{4}{9}\kappa^2\right)^{\frac{1}{2}} b^2}. \quad \text{(A-3)}$$

Finally, equation (A-3) is substituted into eq. (23), i.e., $\langle \cos\theta \rangle_0^{\mathrm{eff}} = kb^2/3$, leading to the following relation as

$$\langle \cos\theta \rangle_0^{\mathrm{eff}} = \frac{2\kappa}{3 + 3\left(1 + \frac{4}{9}\kappa^2\right)^{\frac{1}{2}}}. \quad \text{(A-4)}$$

Equations (A-3) and (A-4) represent that $k$ (and $\langle \cos\theta \rangle_0^{\mathrm{eff}}$) asymptotically approaches either $\frac{3}{b^2}$ (and 1) or $-\frac{3}{b^2}$ (and $-1$) depending on $\kappa$ approaches positive infinity or negative infinity, respectively.

## Appendix. B

In this section, we briefly describe the random phase approximation of microphase separation of block copolymers that obey Gaussian statistics with bending elasticity.

First, we derive the probability distribution of conformation of a single ideal chain without contact interaction between segments. The Hamiltonian of this ideal chain is given by the following matrix form

$$H_0 = \left(\boldsymbol{U}_{1,N-2}\right)^T \boldsymbol{A}_{1,N-2}\boldsymbol{U}_{1,N-2} + \tilde{a}(|\boldsymbol{u}_0|^2 + |\boldsymbol{u}_{N-1}|^2) + \tilde{b}(\boldsymbol{u}_0 \cdot \boldsymbol{u}_1 + \boldsymbol{u}_{N-2} \cdot \boldsymbol{u}_{N-1}), \quad \text{(B-1)}$$

where $\boldsymbol{u}_i \equiv \boldsymbol{r}_{i+1} - \boldsymbol{r}_i$ is the bond vector defined at eq.(7), and $\boldsymbol{U}_{n+1,m-2}$ and $\boldsymbol{A}_{n+1,m-2}$ are defined by

$$\boldsymbol{U}_{1,N-2} = (\boldsymbol{u}_{n+1} \quad \cdots \quad \boldsymbol{u}_i \quad \boldsymbol{u}_{i+1} \quad \cdots \quad \boldsymbol{u}_{m-2}),$$

$$\boldsymbol{A}_{n+1,m-2} = \begin{pmatrix} \boldsymbol{A} & \boldsymbol{B} & 0 & \cdots & \cdots & 0 \\ \boldsymbol{B} & \ddots & \ddots & \ddots & & \vdots \\ 0 & \ddots & \boldsymbol{A} & \boldsymbol{B} & \ddots & \vdots \\ \vdots & \ddots & \boldsymbol{B} & \boldsymbol{A} & \ddots & 0 \\ \vdots & \ddots & \ddots & \ddots & \ddots & \boldsymbol{B} \\ 0 & \cdots & \cdots & 0 & \boldsymbol{B} & \boldsymbol{A} \end{pmatrix} = \begin{pmatrix} \tilde{a}\boldsymbol{I} & \tilde{b}\boldsymbol{I} & 0 & \cdots & \cdots & 0 \\ \tilde{b}\boldsymbol{I} & \ddots & \ddots & \ddots & & \vdots \\ 0 & \ddots & \tilde{a}\boldsymbol{I} & \tilde{b}\boldsymbol{I} & \ddots & \vdots \\ \vdots & \ddots & b\boldsymbol{I} & \tilde{a}\boldsymbol{I} & \ddots & 0 \\ \vdots & \ddots & \ddots & \ddots & \ddots & \tilde{b}\boldsymbol{I} \\ 0 & \cdots & \cdots & 0 & \tilde{b}\boldsymbol{I} & \tilde{a}\boldsymbol{I} \end{pmatrix}, \tag{B-2}$$

where $\boldsymbol{I}$ is the 3-dimensional unit matrix and

$$A = \frac{3k_{\mathrm{B}}T}{2b_0^2}\boldsymbol{I} \equiv \tilde{a}\boldsymbol{I}, \qquad B = -\frac{k_0}{4}\boldsymbol{I} \equiv \tilde{b}\boldsymbol{I}. \tag{B-3}$$

The statistical weight of a conformation of a subchain between $i$-th and $j$-th segments at

$(i, \boldsymbol{r}_i, \boldsymbol{u}_i)$ and $(j, \boldsymbol{r}_j, \boldsymbol{u}_{j-1})$ is given by

$$\begin{aligned} &G_0(i, \boldsymbol{r}_i\boldsymbol{u}_i; j, \boldsymbol{r}_j\boldsymbol{u}_j) \\ &= \int d\boldsymbol{u}_{i+1} \cdots \int d\boldsymbol{u}_{j-2}\, \delta\left(\boldsymbol{r}_j - \boldsymbol{r}_i - (\boldsymbol{u}_i + \cdots + \boldsymbol{u}_{j-1})\right) \\ &\times \exp\left[-\beta\left\{(\boldsymbol{U}_{i+1,j-2})^T \boldsymbol{A}_{i+1,j-2}\boldsymbol{U}_{i+1,j-2} + \tilde{a}\left(|\boldsymbol{u}_i|^2 + |\boldsymbol{u}_{j-1}|^2\right) + \tilde{b}(\boldsymbol{u}_i \cdot \boldsymbol{u}_{i+1} + \boldsymbol{u}_{j-2} \cdot \boldsymbol{u}_{j-1})\right\}\right]. \end{aligned} \tag{B-4}$$

Using the Fourier transform, we obtain

$$\delta\left(\boldsymbol{r}_j - \boldsymbol{r}_i - (\boldsymbol{u}_i + \cdots + \boldsymbol{u}_{j-1})\right) = \frac{1}{(2\pi)^3}\int d\boldsymbol{q}\, \exp[i\boldsymbol{q} \cdot \{(\boldsymbol{r}_j - \boldsymbol{r}_i) - (\boldsymbol{u}_i + \cdots + \boldsymbol{u}_{j-1})\}], \tag{B-5}$$

Equation (B-4) can be rewritten as

$$\begin{aligned} &G_0(i, \boldsymbol{r}_i\boldsymbol{u}_i; j, \boldsymbol{r}_j\boldsymbol{u}_j) \\ &= \frac{1}{(2\pi)^3}\exp\left[-\beta a\left(|\boldsymbol{u}_i|^2 + |\boldsymbol{u}_{j-1}|^2\right)\right]\int d\boldsymbol{q}\, \exp[i\boldsymbol{q} \cdot \{(\boldsymbol{r}_j - \boldsymbol{r}_i) - (\boldsymbol{u}_i + \boldsymbol{u}_{j-1})\}] \\ &\times \int d\boldsymbol{u}_{i+1} \cdots \int d\boldsymbol{u}_{j-2}\, \exp\left[-\beta\left\{\widetilde{\boldsymbol{U}}_{i+1,j-2}^T \boldsymbol{A}_{i+1,j-2}\widetilde{\boldsymbol{U}}_{i+1,j-2} - \frac{1}{4}\boldsymbol{B}_{i+1,j-2}^T \boldsymbol{A}_{i+1,j-2}^{-1}\boldsymbol{B}_{i+1,j-2}\right\}\right], \end{aligned} \tag{B-6}$$

where we defined

$$\boldsymbol{B}_{i+1,j-2}(\boldsymbol{u}_i, \boldsymbol{u}_{j-1}) = \begin{pmatrix} b\boldsymbol{u}_i + \frac{i}{\beta}\boldsymbol{q} \\ \frac{i}{\beta}\boldsymbol{q} \\ \vdots \\ \frac{i}{\beta}\boldsymbol{q} \\ b\boldsymbol{u}_{j-1} + \frac{i}{\beta}\boldsymbol{q} \end{pmatrix} \tag{B-7}$$

and

$$\widetilde{\boldsymbol{U}}_{i+1,j-2}^T \equiv \boldsymbol{U}_{i+1,j-2}^T + \frac{1}{2}\boldsymbol{A}_{i+1,j-2}^{-1}\boldsymbol{B}_{i+1,j-2}. \tag{B-8}$$

As the eigenvalues of the matrix

$$\boldsymbol{A}_{n+1,m-2} = \begin{pmatrix} \boldsymbol{A} & \boldsymbol{B} & 0 & \cdots & \cdots & 0 \\ \boldsymbol{B} & \ddots & \ddots & \ddots & & \vdots \\ 0 & \ddots & \boldsymbol{A} & \boldsymbol{B} & \ddots & \vdots \\ \vdots & \ddots & \boldsymbol{B} & \boldsymbol{A} & \ddots & 0 \\ \vdots & \ddots & \ddots & \ddots & \ddots & \boldsymbol{B} \\ 0 & \cdots & \cdots & 0 & \boldsymbol{B} & \boldsymbol{A} \end{pmatrix}, \tag{B-9}$$

are given by

$$\lambda_k \equiv A + 2B\cos\left(\frac{k\pi}{(j-2)-(i+1)+1}\right) \quad (k: i+1, \cdots, j-2), \tag{B-10}$$

the Gaussian integral in eq.(B-6) can be performed to give

$$\begin{aligned} &G_0(i, \boldsymbol{r}_i\boldsymbol{u}_i; j, \boldsymbol{r}_j\boldsymbol{u}_j) \\ &= \frac{1}{(2\pi)^3}\exp\left[-\beta a\left(|\boldsymbol{u}_i|^2 + |\boldsymbol{u}_{j-1}|^2\right)\right]\int d\boldsymbol{q}\exp[i\boldsymbol{q}\cdot\{(\boldsymbol{r}_j - \boldsymbol{r}_i) - (\boldsymbol{u}_i + \boldsymbol{u}_{j-1})\}] \\ &\quad \times \exp\left[\frac{\beta}{4}\boldsymbol{B}_{i+1,j-2}^T(\boldsymbol{A}_{i+1,j-2})^{-1}\boldsymbol{B}_{i+1,j-2}\right] \times \prod_{k=i+1}^{j-2}\left(\frac{\pi}{\beta\lambda_k}\right)^{\frac{3}{2}}. \end{aligned} \tag{B-11}$$

with eq. (B-11). Then, we obtain the probability $P(n, \boldsymbol{r}; m, \boldsymbol{r}')$ of finding *n*-th and *m*-th segments at ***r*** and ***r’*** as

$$
\begin{aligned}
&P(n,\boldsymbol{r};m,\boldsymbol{r}')\\
&\quad=\frac{1}{\mathfrak{N}}\int d\boldsymbol{r}_0\int d\boldsymbol{u}_0\int d\boldsymbol{r}_N\int d\boldsymbol{u}_{N-1}\int d\boldsymbol{u}_{n-1}\int d\boldsymbol{u}_n\int d\boldsymbol{u}_{m-1}\int d\boldsymbol{u}_m\\
&\qquad\times G_0(0,\boldsymbol{r}_0,\boldsymbol{u}_0;n,\boldsymbol{r},\boldsymbol{u}_{n-1})\ \exp\left[-\beta\left(-\frac{k_0}{2}\boldsymbol{u}_{n-1}\cdot\boldsymbol{u}_n\right)\right]\\
&\qquad\times G_0(0,\boldsymbol{r},\boldsymbol{u}_n;m,\boldsymbol{r}',\boldsymbol{u}_{m-1})\ \exp\left[-\beta\left(-\frac{k_0}{2}\boldsymbol{u}_{m-1}\cdot\boldsymbol{u}_m\right)\right]\\
&\qquad\times G_0(m,\boldsymbol{r}',\boldsymbol{u}_m;N,\boldsymbol{u}_N,\boldsymbol{u}_{N-1}),
\end{aligned}
\tag{B-12}
$$

where $\mathfrak{N}$ is the normalization factor.

Using the fact that the tridiagonal matrix of rank $n$

$$
\widetilde{\boldsymbol{A}}_n\equiv\begin{pmatrix}\tilde{a} & \tilde{b} & 0 & \cdots & 0\\ \tilde{b} & \tilde{a} & \ddots & \ddots & \vdots\\ 0 & \ddots & \ddots & \ddots & 0\\ \vdots & \ddots & \ddots & \tilde{a} & \tilde{b}\\ 0 & \dots & 0 & \tilde{b} & \tilde{a}\end{pmatrix},
\tag{B-13}
$$

has its inverse matrix with the $(k,m)$-element

$$
\left(\widetilde{\boldsymbol{A}}_n^{-1}\right)_{k,m}=\frac{1}{n+1}\sum_{l=1}^{n}\sin\left(\frac{\pi}{n+1}kl\right)\sin\left(\frac{\pi}{n+1}ml\right)\frac{1}{\tilde{a}+2\tilde{b}\cos\frac{\pi}{n+1}l},
\tag{B-14}
$$

we can perform the integrations in eq. (B-12).

After a rather lengthy calculation, we finally obtain

$$
P(n,\boldsymbol{r};m,\boldsymbol{r}')=const.\times\exp\left[-\frac{|\boldsymbol{r}'-\boldsymbol{r}|^2}{2\Pi_{n,m-1}}\right],
\tag{B-15}
$$

where we defined

$$
\begin{aligned}
&\Pi_{n,m-1}\\
&=\frac{1}{\beta(m-n-1)}\left[\sum_{l=1}^{m-n-2}\frac{1}{\tilde{a}+2\tilde{b}\cos\frac{\pi}{m-n-1}l}\left[\frac{1-(-1)^l}{2}\frac{1+\cos\left\{\frac{\pi}{m-n-1}l\right\}}{\sin\left\{\frac{\pi}{m-n-1}l\right\}}\right]^2\right].
\end{aligned}
\tag{B-16}
$$

For Table of Contents Only

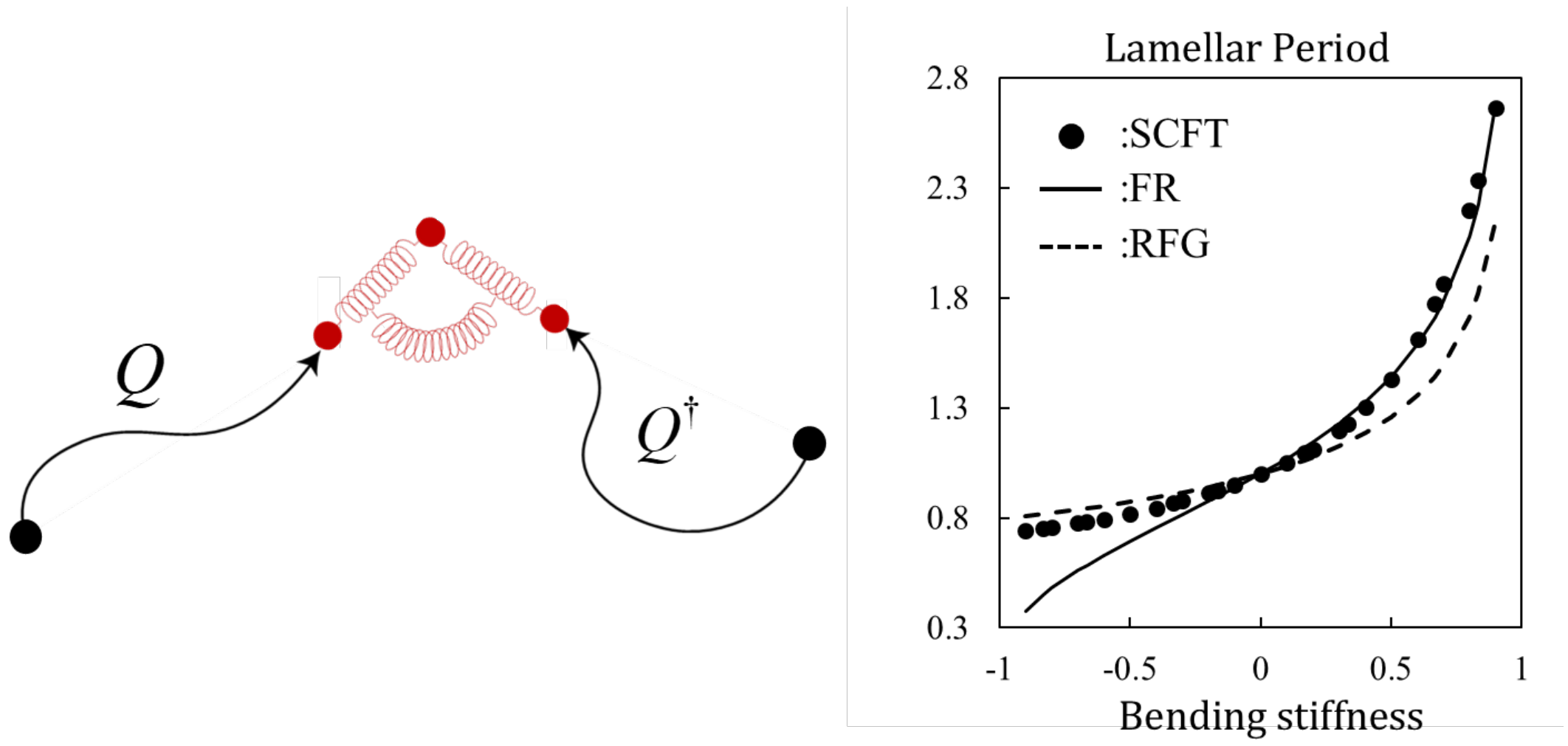